\documentclass[rmp,graphics,onecolumn,aps,floats,epsfig]{revtex4}

\usepackage{graphicx}
\usepackage{psfrag}

\begin{document}

 \title{Information and fundamental elements of the structure of quantum theory}
 \author{\v Caslav Brukner and Anton Zeilinger \\
 Institut f\"ur Experimentalphysik, Universit\"at Wien,\\
 Boltzmanngasse 5, A--1090 Wien, Austria}

 \date{\today}


 \begin{abstract}

 Niels Bohr wrote: ''There is no quantum world. There is only an
 abstract quantum physical description. It is wrong to think that
 the task of physics is to find out how Nature is. Physics
 concerns what we can say about Nature.'' In an analogous way, von
 Weizs{\"a}cker suggested that the notion of the elementary alternative, the "Ur", should
 play a pivotal role when constructing physics. Both approaches suggest that the
 concept of information should play an essential role in the foundations of any
 scientific description of Nature. We show that if, in our
 description of Nature, we use one definite proposition per
 elementary constituent of Nature, some of the essential
 characteristics of quantum physics, such as the irreducible randomness of
 individual events, quantum complementary and quantum entanglement,
 arise in a natural way. Then quantum physics is an elementary theory of
 information.

 \end{abstract}

\maketitle

 {\bf Dedicated to Prof. C. F. von Weizs\"{a}cker at the occasion
 of his $\bf{90^{th}}$ birthday.}


 \section{Introduction}

 All our description of objects is represented by propositions. The
 use of propositions is not a matter of our choice. In contrast,
 it is a necessity which is behind each of our attempts to
 learn something new about Nature and to communicate this
 knowledge with others. It is a necessity which we follow
 constantly and without any intention and it seems that there is
 no way to avoid it even if the phenomena to be described and to
 be understood are highly counterintuitive and distinct from both
 our everyday experience and the classical world view. One may
 even say that there is no need to avoid it. The reason is that the
 only way we are able to understand any phenomena in Nature,
 including quantum phenomena, is exclusively through the
 epistemological structure of classical physics and everyday
 experience. Bohr (1949) emphasized that ''How far the [quantum]
 phenomena transcend the scope of classical physical explanation,
 the account of all evidence must be expressed in classical terms.
 The argument is simply that by the word 'experiment' we refer
 to a situation where we can tell others what we have done and
 what we have learned and that, therefore, the account of the
 experimental arrangement and the result of observation must be
 expressed in unambiguous language with suitable application of
 the terminology of classical physics.''. von Weizs\"{a}cker (1974) emphasized that
 the understanding of new physical theories will also be given
 in language: "This verbalized language must be the language spoken by
 those physicists who do not know yet the theory we are telling them. The language used in
 order to explain a theory which we propose in mathematical form
 is the language which has been existed before the theory. On the
 other hand, it is not self-evident that this language has a clear
 meaning at all, because if it had a completely clear meaning
 probable the new theory would not be needed. Thus it may happen
 that by applying our new formalism to experience - an application
 made possible by our existing language - we may tacitly or
 explicitly change the rules of this very language.''

 Rigorously speaking a system is nothing else than a construct based on
 a complete list of propositions together with their truth values.
 The propositions from the list could be (1) ''The
 velocity of the object is $v$'' or (2) ''The position of the
 object is $x$'' and could be associated both to classical and to
 quantum objects. Yet, there is an important difference between the two cases.
 From the theorems of Bell (1964) and of
 Kochen and Specker (1967) we know that for a quantum system one
 cannot assert definite (noncontextual) truth values to all conceivable
 propositions simultaneously. For example, if the proposition (1) above
 is a definite proposition, then the proposition (2) must necessarily be
 completely indefinite and vice versa. The two propositions are
 mutually exclusive. This is a specific case of quantum complementarity.

 Therefore, in an attempt to describe quantum phenomena we are
 unavoidably put in the following situation. On one hand the
 epistemological structure applied has to be inherited from the
 classical physics: the description of a quantum system has to be
 represented by the propositions which are used in the description
 of a classical system, and on the other hand, those propositions
 cannot be assigned to a quantum system
 simultaneously. Now, a natural question arises: How to join these two, seemingly
 inconsistent, requirements? We suggest to use the concept of
 "knowledge" or "information". Then even in situations where
 we cannot assert
 simultaneously definite truth-values to mutually exclusive
 propositions we can assert measures of information about
 their truth values. The structure of the theory including the
 description of the time evolution can then be expressed in terms
 of measures of information\footnote{Heisenberg (1958) wrote:
 ''The laws of nature which we formulate mathematically in quantum
 theory deal no longer with the particles themselves but with our
 knowledge of the elementary particles. ... The conception of
 objective reality ... evaporated into the ... mathematics that
 represents no longer the behavior of elementary particles but
 rather {\em our knowledge of this behavior}.''}. To us this seems
 to be a change with the lowest possible "costs" in the
 epistemological structure of classical physics. And since some
 costs are unavoidable anyway we believe
 that the information-theoretical formulation of quantum physics leads
 to the ''easiest" understanding of the theory.

 From the point of view that the information content of a quantum system is
 fundamentally limited we will discuss precisely the empirical
 significance of the terms involved in formulating quantum theory,
 particularly the notion of a quantum state. However we are aware
 of the possibility that this might not carry the same degree of
 intuitive appeal for everyone. It is clear that it may be matter
 of taste whether one accepts the suggested concepts and principles as
 self-evident as we do or not. If not, then one may turn the reasoning
 around and, following our approach in (Brukner and Zeilinger, 1999;
 Brukner {\it et al.} 2001), argue for the validity of the statements
 given in the paper on the basis of known features of
quantum physics.

 The conceptual groundwork for the ideas presented here has been
 prepared most notably by von Bohr (1958), Weizs\"{a}cker (1958)
 and Wheeler (1983). In contrast to those other authors who look for
 deterministic mechanisms hidden behind the observed facts, these
 authors attempt to understand the structure of quantum theory as
 a necessity for extracting  whatever meaning from the data of
 observations.

 In recent years several different ideas were put forward
 suggesting that information can help us to learn more about the
 foundations of quantum physics. The foundations of quantum
 mechanics are interpreted in the light of quantum information
 (Fuchs, 2001; 2002, Caves {\it et al.}, 2001a; 2001b). It was also suggested
 how to reduce quantum theory to few statements of physical
 significance by generalizing and extending classical
 probability theory (Hardy, 2001a; 2001b). In another approach it was shown how
 certain elements of the structure of quantum theory emerge from looking for
 invariants of probabilistic observations assuming that any newly gained
 information shall lead to more accurate knowledge of these
 invariants (Summhammer, 1988; 1994; 2000; 2001).

 \section{Finiteness of Information, Ur, Elementary System}

 One of the most distinct features of quantum physics with respect
 to classical physics is that prediction with certainty of individual
 outcomes is only possible for a very limited class
 of experiments. Such a prediction is equivalent to saying
 that the corresponding propositions have definite truth values. For all
 other (complementary) propositions the truth values are necessarily
 indefinite. We suggest this to be a consequence of the feature
 that\footnote{Feynman wrote: "It  always bothers me that, according to the laws as we understand
 them today, it takes a computing machine an infinite number of
 logical operations to figure out what goes on in no matter how
 tiny a region of space and no matter how tiny a region of time,
 ... why should it take an infinite amount of logic to figure out
 what one tiny piece of space-time is going to do?" A closely related view was assumed by Landauer, who writes
 in his article ''Information is Physical'' (1991):
''... the laws of physics are ... limited by the range of
information processing available.''.}

 {\it The information content of a quantum system is finite.}

 With this we mean that a quantum system cannot carry enough
 information to provide definite answers to all questions that
 could be asked experimentally. Then, by necessity the answer of the
 quantum system to some questions must contain an element of randomness.
 This kind of randomness must then be irreducible, that is, it
 cannot be reduced to ''hidden'' properties of the system.
 Otherwise the system would carry more information than what is
 available. Thus, without any additional physical structure
 assumed, we let the irreducible randomness of an individual event
 and complementarity, be a consequence of the finiteness of
 information.

 How much information is available to a quantum system? If this
 information is limited than it is
 natural to assume that if we decompose a physical system, which
 may be represented by numerous propositions, into its constituents,
 each such constituent will be described by fewer
 propositions. This process of subdividing a system can go further
 until we reach a final limit when an individual system represents
 the truth value to one single proposition only. It is then
 suggestive to replace the above statement by a more precise one
 (Zeilinger, 1999):

 {\it The most elementary system represents the truth value of one
 proposition.}

 We call this the principle of quantization of information. One may
 consider the above statement as a definition of what is the most
 elementary system. Note that the truth value of a proposition can
 be represented by one bit of information with ''true'' being
 identified with the bit value ''1'' and ''false'' being identified
 with the bit value ''0''. Thus, the principle becomes simply:

 {\it The most elementary system carries 1 bit of information.}

 We relate the notion of the most elementary system to that of the
 "Ur" introduced by von Weizs\"{a}cker. He was the first who
 introduced the concept of the most basic informational constituent
 of all objects (''Ur''). von Weizs\"{a}cker wrote (1974): ''It is
 certainly possible to decide any large alternative step by step
 in binary alternatives. This may tempt us to describe all objects
 as composite systems composed from the most simple possible
 objects. The simplest possible object is an object with a
 two-dimensional Hilbert space, the 'ur'. The word 'ur' is
 introduced to have an abstract term for something which can be
 described by quantum theory and has a two-dimensional Hilbert
 space, and nothing more.''

 \begin{figure}
 \includegraphics[angle=0,width=0.3\textwidth]{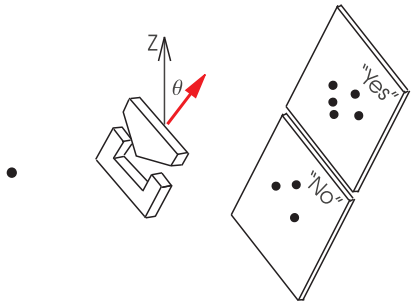} \caption{Spin
 measurement of a spin-1/2 particle. The particle passes through
 the Stern-Gerlach magnet oriented at the angle $\theta$, and then
 it hits one of the detector plates behind the Stern-Gerlach
 magnet. Depending on whether the upper or the lower detector plate
 is hit by a particle we call the outcome ''yes'' and ''no'',
 respectively.} \label{plates}
 \end{figure}

 How much information is contained in more complex systems
 consisting of $N$ elementary systems? It is natural to assume that the
 information content of a complex system is proportional to the
 number of elementary constituents. The principle of
 quantization of information is thus generalized to (Zeilinger, 1999)

 {\it N elementary systems represent the truth values of N
 propositions,}

 or equivalently,

 {\it N elementary systems carry N bits.}

 The finding compatible with this principle were reported (Donath and Svozil, 2002; Svozil 2002).
 Again one may consider the two statements as definitions of what is a composite system
 consisting of $N$ elementary systems. Note that the principle given above does not make any statement about how the
 information contained in $N$
 propositions ($N$ bits) is distributed over the $N$ systems. It can be
 represented by the $N$ systems individually or, alternatively, it can be
 represented by $N$ systems jointly. The latter is the feature of
 quantum entanglement, discussed in more detail below, for which
 Schr\"{o}dinger (1935) wrote:
''If two separated bodies, each by itself known maximally, enter
a situation in which they influence each other, and separate
again, then there occurs regularly that which I have just called
entanglement of our knowledge of the two bodies. The combined
expectation-catalog consists initially of a logical sum of the
individual catalogs; during the process it develops causally in
accord with known law (there is no question whatever of
measurement here). The knowledge remains maximal, but at its end,
if the two bodies have again separated, it is not again split
into a logical sum of knowledges about the individual bodies.
What still remains of that may have becomes less than maximal,
even very strongly so. -One notes the great difference over
against the classical model theory, where of course from known
initial states and with known interaction the individual end
states would be exactly known.''

 The fundamental statements given above seem to suggest that binary (yes-no)
 alternatives are representatives of basic information units of all systems.
 Consider the case of  $n$-fold (i.e. ternary, quanternany etc.) alternatives
 generalizing the binary ones. Obviously, any $n\!=\!2^N$-fold alternative is decomposable
 into binary ones. Note that such an alternative can be realized in measurement of
 $N$ elementary systems. Yet, it is not obvious how to decompose or how to consider a
 general $n$-fold alternative with $n \!\neq \!2^N$? An interesting possibility would be to
 find the factorization of number $n$ into its prime-number factors $p_1, p_2, ...$  and then
 to decompose the $n$-fold alternative into a sequential serial of $p_1$-fold  alternatives, $p_2$-fold
 alternatives etc. Obviously such an approach would require to extend the notion of elementary system
 to all prime-number dimensional systems. Interestingly, as it will be shown later
 (see Sec. \ref{brot}), only in these cases where the dimension of the quantum system is
 (a power of) a prime number, the total information content of the system can be defined
 unambiguously. This might suggest that the notion of the elementary system should indeed be extended to all
 prime-number dimensions. However,
 in this manuscript we will mainly restrict our analysis to binary decomposable alternatives.

 We would like to stress again that notions such as that a system
 ''represents'' the truth value of a proposition or that it
 ''carries'' one bit of information only implies a statement concerning
 what can be said about possible measurement results. For us a system is no more
 than a representative of a proposition.

 \section{Mutually complementary propositions}
 \label{NNumber}

 We consider an explicit example of an elementary system, the spin-1/2
 particle, and the Stern-Gerlach experiment as depicted in Fig.
 \ref{plates} schematically. Depending on whether the upper or the
 lower detector plate is hit by a particle we call the outcome
 ''yes'' and ''no'' respectively, where "yes" and "no"
 represent the truth values of the proposition for the spin to be up
 along a chosen direction. The upper detector plate is hit
 with probability $p$. If it is not hit the other detector plate
 will be hit with probability $1-p$. Therefore we consider a binary
 alternative. Different experimental situations are specified by
 the orientation $\theta$ of the magnet in the Stern-Gerlach
 apparatus as shown in Fig. \ref{plates}.

\begin{figure}
 \includegraphics[angle=0,width=0.65\textwidth]{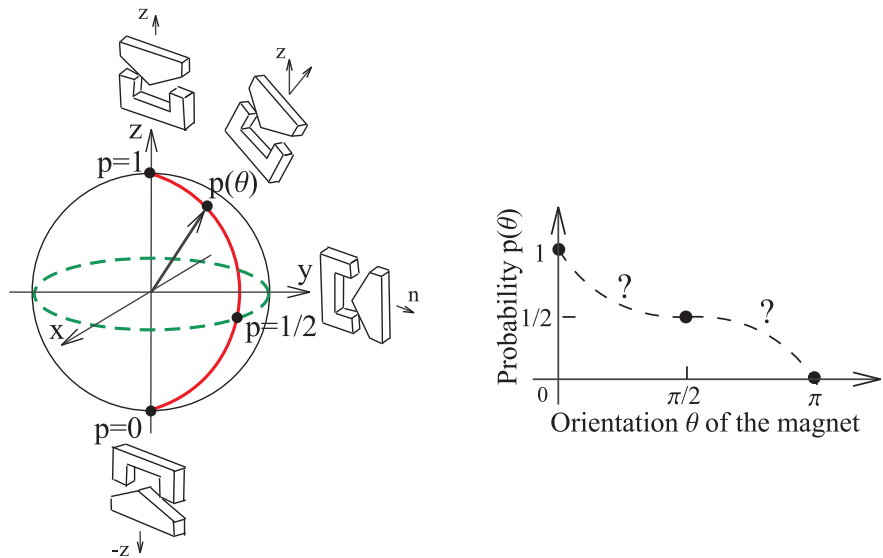} \caption{The
 gradual change of the probability $p(\theta)$ of a ''yes'' (''spin
 up'') count with a gradual change of the orientation $\theta$ of
 the magnet in the Stern-Gerlach apparatus. The measurement along
 the z-axis gives the result ''yes'' with certainty. Because of the
 symmetry of the problem the probabilities for a ''yes'' and for a ''no''
 count in a measurement along any direction in the $x$-$y$ plane
 (the green circle) are equal (=1/2). How does the probability
 $p(\theta)$ of a ''yes'' count depend on $\theta$ explicitly?}
 \label{kontinum}
 \end{figure}

 Consider an elementary system specified by the true proposition ''The spin
 along the $z$-axis is up'' (or, equivalently,
 by the false proposition ''The spin along
 the $-z$-axis is up''). This situation is described by the
 probabilities $p(0)\!=\!1$ and $p(\pi)\!=\!0$ for the ''yes''
 outcome. Because a spin can carry one bit of information only, each proposition: ''The spin along the
 direction tilted at an angle $\theta$ $(0 \!<\! \theta\! <\! \pi)$ from the
 $z$-axes is up'' has to be probabilistic (Fig. \ref{kontinum}). How does
 the probability $p(\theta)$ of a ''yes'' count depend upon the
 angle $\theta$?

\begin{figure}
 \includegraphics[angle=0,width=0.7\textwidth]{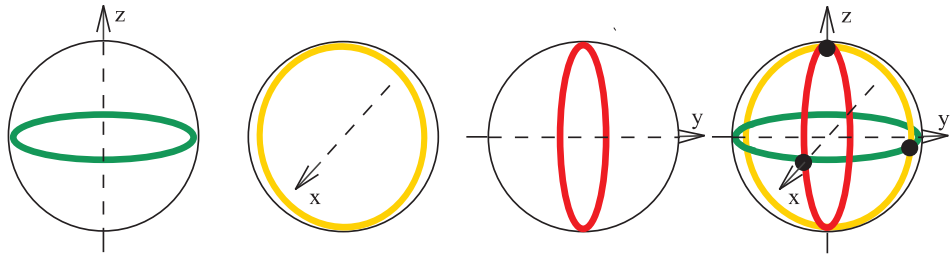}
 \caption{The formation of mutually complementary propositions
 associated with orthogonal spin components. If measurement along
 the $z$-axis ($x$-axis) [$y$-axis] gives a definite result, measurement
 along any direction in the $x$-$y$ plane, the green circle ($y$-$z$ plane,
 the yellow circle) or [$x$-$z$ plane, the red circle] will be
 maximally random, respectively. There are altogether three
 mutually complementary spin measurements represented by three
 intersection points of the green, yellow, and red circle.}
 \label{spin}
 \end{figure}

 We assume that the mapping of $\theta$ to $p(\theta)$ is
 analytic\footnote{If, in contrast, $p(\theta)$ would only be
 sectionally analytic in $\theta$ then there would be points of
 nonanalyticity separating two regions in which the function
 $p(\theta)$ has different analytic forms. Thus the values of the
 function on a finite segment in the interior of a domain of
 analyticity would only determine, by the uniqueness theorem for
 analytic functions, the function up to the next point of
 nonanalyticity. Clearly, to describe such a system completely we
 would need catalogs both of functional values on finite segments
 in the interior of each domain of analyticity and of the positions
 of the points of nonanalyticity. Such a catalog would require
 large amount of information to describe the functional
 dependence and thus contradicts our desideratum of minimal information
 content of a quantum system.} and monotonic. Then using the Cauchy theorem
 about continuous and monotonic functions one concludes that there
 has to be one and only one angle of orientation of the magnet in
 the Stern-Gerlach apparatus where the probabilities for a ''yes''
 and for a ''no'' outcome are equal. Because of the symmetry of
 the problem this obviously has to be the angle $\pi/2$. For each
 direction $\vec{n}$ in the $x$-$y$ plane (the green circle on the
 sphere in Fig. \ref{kontinum} and \ref{spin}) the proposition
 ''The spin along the $\vec{n}$-axis is up'' is completely
 indefinite, that is, we have absolutely no knowledge which
 outcome ''yes'' or ''no'' will be observed in a specific
 individual measurement.

 Note, however, that in principle this equal number of yes-no
 outcomes could also be achieved by an ensemble of systems each
 giving a definite result for each direction such that the same
 number of ''yes'' or ''no'' results is obtained. Yet again this
 would imply that an individual system carries enough information
 to permit assignment of definite truth values to all possible
 propositions, in contradiction to our basic principle.

 Consider now the state of a spin-1/2 particle specified by the
 proposition ''The spin along the $x$-axis is up (down)''. In this
 case we have complete knowledge which outcome will be observed
 when the Stern-Gerlach magnet is oriented along the $\pm x$-axis at
 the expense of the fact that we have absolutely no knowledge about
 the outcome for the orientation of the magnet along \it any \rm
 direction in the $y$-$z$ plane (the yellow circle on the sphere in
 Fig. \ref{spin}).

 Finally, consider the state of a spin-1/2 particle specified by
 the proposition ''The spin along the $y$-axis is up (down)''. In
 that case we know precisely the outcome of the experiment when the
 Stern-Gerlach magnet is oriented along the $\pm y$-axis at the
 expense of complete uncertainty about the outcome when it is
 oriented along any direction in the $x-z$ plane (the red circle on
 the sphere in Fig. \ref{spin}).

 There are, therefore, altogether {\it three mutually exclusive} or
 {\it complementary} propositions (represented by three
 intersection points of the green, yellow and red circle on the
 sphere in Fig. \ref{spin}): ''The spin along direction $\vec{n}_1$
 is up (down)'', ''The spin along direction $\vec{n}_2$ is up
 (down)'' and ''The spin along direction $\vec{n}_3$ is up (down)'',
 where $\vec{n}_1$, $\vec{n}_2$ and $\vec{n}_3$ are mutually
 orthogonal directions. These are propositions with a property of
 mutually exclusiveness: the total knowledge of one proposition is
 only possible at the cost of total ignorance about the other two
 complementary ones. In other words precise knowledge of the outcome of one
 experiment implies that all possible outcomes of
 complementary ones are equally probable.

 Why are there exactly 3 mutually complementary propositions for
 the elementary system and not, e.g., 2 or 4? We do not
 understand that fully. However the discussion above indicates that
 there is a strong link between the number (3) of mutually
 complementary propositions and the (three-)dimensionality of the ordinary
 space. We will come back once more to these question in the conclusions.
 But it is important to note that in any system with dichotomic (2-valued)
 observables there are always three complementary propositions even if these
 cannot be linked to the dimensionality of ordinary space.

 \section{Measure of information in a probabilistic experiment}

 Consider a probabilistic experiment with $n$ possible outcomes.
 Suppose that the experimenter plans to perform $N$ trials of the
 experiment. All he knows before the trials are performed are the probabilities
 $p_1, ... p_i, ..., p_n$ for all possible outcomes to occur:
 What kind of prediction can the experimenter make?

 In general two cases are conceivable. The experimenter can ask: "What is
 the precise  sequence of the $N$ outcomes?" or "What is the number of occurrences of the
 outcome $i$?". We will say that in answering the first question the experimenter makes a
 "deterministic" prediction and in answering the second one he makes a "probabilistic"
 prediction\footnote{Summhammer (2000) wrote: ''I want to discard a deterministic link.
The reason is that the amount of records available to the observer to form a conception of the 
world is always finite, so that many different sets of laws can be invented to account for 
them. Pinning down any one of these sets as {\it the} laws of nature is then purely 
speculative. On the other hand we have the probabilistic view,
which is successfully used to interpret quantum observations. It
seems that in this view we assign a minimum of information
content to observed data. To see this, imagine the $N$ trials of
a probabilistic yes-no experiment, like tossing a coin, in which
the outcome "yes" occurs $L$ times. If we want to tell somebody
else the result it is sufficient to state the values of $N$ and
of $L$. With the deterministic view, in which the precise
sequence of outcomes is important, we would in general have to
communicate many more details to enable the receiver to
reconstruct this sequence.}
 (following discussion in Summhammer, 2000). Obviously the deterministic prediction can only
 make sense if different outcomes follow from the intrinsically different
 individuals of the ensemble measured - the situation which we have in classical measurements. Then the precise
 sequence of outcomes reveal which property which individual member of the ensemble
 carries. One
 can show that for sufficiently large $N$ Shannon's measure of information (Shannon, 1948)
 \begin{equation}
 H=-\sum_{i=1}^{n} p_i \log p_i \label{mrgod}
 \end{equation}
 is equal to the mean minimal number
 of yes-no questions (when the logarithm in Eq. (\ref{mrgod}) is
 taken to base 2) necessary to determine which particular sequence
 of outcomes occurs, divided by $N$. As we show in more detail elsewhere (Brukner
 and Zeilinger, 2001) this suggests that Shannon's measure of  information is the adequate
 measure of the uncertainty in the deterministic prediction and and thus in classical measurements.

 In contrast to classical measurements, quantum measurements, in general,
 cannot be claimed to reveal a property of the individual quantum
 system existing before the measurement is performed. This causes certain
 conceptual difficulties when we try to define the information gain in
 quantum measurements using the notion of Shannon's measure (for discussion
 of these points see Brukner and Zeilinger, 2001; Hall, 2000; Brukner and
 Zeilinger 2000; Timpson 2001). Since outcomes of quantum-mechanical
 experiments are in general intrinsically  probabilistic, there the experimenter
 can only make probabilistic predictions. If the experimenter decides to perform
 $N$ future experimental trails, all he can guess is the number of occurrences
 of a specific outcome. Such a prediction will now be analyzed for the case of
 two possible outcomes "yes" and "no".

 Because of the statistical fluctuations
 associated with any finite number of experimental trials, the
 number $L$ of occurrences of the "yes" outcome in future
 repetitions of the experiment is not precisely
 predictable\footnote{Here, a very subtle and careful position was assumed by
 Weizs\"acker (1974) who writes: ''It is most important to see that
 this [the fact that probability is not a prediction of the
 precise value of the relative frequency] is not a particular
 weakness of the objective empirical use of the concept of
 probability, but a feature of the objective empirical use of any
 quantitative concept. If you predict that some physical quantity,
 say a temperature, will have a certain value when measured, this
 prediction also means its expectation value within a statistical
 ensemble of measurements. The same statement applies to the
 empirical quantity called relative frequency. But here are two
 differences which are connected to each other. The first
 difference: In other empirical quantities the dispersion of the
 distribution is in most cases an independent empirical property
 of the distribution and can be al.tered by more precise
 measurements of other devices; in probability the dispersion is
 derived from the theory itself and depends on the absolute number
 of cases. The second difference: In other empirical quantities
 the discussion of their statistical distributions is done by
 another theory than the one to which they individually belong,
 namely by the general theory of probability; in probability this
 discussion evidently belongs to the theory of this quantity,
 namely of probability itself. The second difference explains the
 first one.''}. The random variable $L$ is subject to a binomial
 distribution. Since it has a finite $\sigma$ deviation, it fulfills
 Chebyshev's inequality (Gnedenko, 1976):
 \begin{equation}
 \mbox{Prob}\{|L-pN| > k \sigma \} \leq \frac{1}{k^2},
 \label{nejednacina}
 \end{equation}
 where the standard deviation $\sigma$ is given by
 \begin{equation}
 \sigma= \sqrt{{p(1-p)} {N}}. \label{error}
 \end{equation}
 This inequality means that the probability that the number $L$ will
 deviate from the product $pN$ by more often than $k$ deviations is less
 than or equal to $1/k^2$. In the case of small $\sigma$, large
 deviations of the number of occurrences of the ''yes'' outcome
 from the mean value $pN$ are improbable. In this case the experimenter
 knows the future number of occurrences with a high certainty.
 Conversely, a large $\sigma$ indicates that not all highly
 probable values of $L$ lie near the mean $pN$. In that case
 the experimenter knows much less about the future number of
 occurrences.

 We suggest to identify the experimenter's uncertainty $U$ with
 $\sigma^2$. Then it will be proportional to the number of trials.
 This important property guarantees that each individual
 performance of the experiment contributes the same amount of
 information, no matter how many times the experiment has already
 been performed. After each trial the experimenter's uncertainty
 about the specific outcome therefore decreases by
 \begin{equation}
 U = \frac{\sigma^2}{N}=p(1-p).
 \end{equation}
 This is the lack of information about a specific outcome  with
 respect to a single future experimental trial. If, instead of two
 outcomes, we have $n$ of them with the probabilities $\vec{p}
 \equiv (p_1, p_2, ... p_n)$ for the individual occurrences, then we suggest to
 define the total lack of information regarding all $n$ possible
 experimental outcomes as
 \begin{equation}
 U(\vec{p}) =\sum_{j=1}^{n} U(p_j) = \sum_{j=1}^{n}p_j(1-p_j)=1-\sum_{j=1}^{n}
 p^2_j.
 \end{equation}
 The uncertainty is minimal if one probability is equal to one and
 it is maximal if all probabilities are equal.

 This suggests that the knowledge, or information, with respect to
 a single future experimental trial an experimentalist possesses
 before the experiment is performed is somehow the complement of
 $U(\vec{p})$ and, furthermore, that it is a function of a sum of
 the squares of probabilities. A first ansatz therefore would be
 $I(\vec{p}) = 1 - U(\vec{p}) = \sum_{i=1}^{n} p^2_i$. Expressions
 of such a general type were studied in detail by Hardy, Littlewood
 and P\'{o}lya (1952). Notice that this expression can also be viewed
 as describing the length of the probability vector $\vec{p}$.
 Obviously, because of $\sum_{i} p_i=1$, not all vectors in
 probability space are possible. Indeed, the minimum length of
 $\vec{p}$ is given when all $p_i$ are equal ($p_i\!=\!1/n$). This
 corresponds to the situation of complete lack of information in an
 experiment about its future outcome. Therefore we suggest to normalize the
 measure of information in an individual quantum measurement as obtaining
 finally
 \begin{equation}
 I(\vec{p})=  {\cal N} \sum_{i=1}^{n} \left( p_i-\frac{1}{n}
 \right)^2, \label{junko}
 \end{equation}
 where $\cal{N}$ is the normalization\footnote{In (Brukner and
 Zeilinger, 1999) only those cases were considered where maximally
 $k$ bits of information can be encoded, i.e. $n=2^k$. The
 normalization there is ${\cal N} = 2^k k/(2^k-1)$. Then $I(\vec{p})$
 results in $k$ bits of information if one $p_i=1$ and it results
 in 0 bits of information when all $p_i$ are equal.}. Specifically,
 for a binary experiment the measure of information is given as
 \begin{equation}
 I(p_1,p_2)=  2 \left(p_1-\frac{1}{2} \right)^2 + 2 \left(
 p_2-\frac{1}{2} \right)^2 =(p_1-p_2)^2. \label{sunka}
 \end{equation}
 It reaches its maximal value of 1 bit of information if one of the probabilities
 is one and it takes its minimal value of 0 bits of information if both
 probabilities are equal.

 \section{The catalog of knowledge of a quantum system}

 Consider again a stationary experimental arrangement with two
 detectors, where only one detector fires in each experimental
 trial. The first detector, say, fires (we call this the ''yes'' outcome)
 with probability $p_1$. If it is does not fire the other detector
 fires with probability $p_2=1-p_1$ (the ''no'' outcome).

 Note that the experimenter's measure of information for the binary
 experiment as defined by Eq. (\ref{sunka}) is invariant under permutation of
 the set of possible outcomes. In other words, it is a symmetrical
 function of $p_1$ and $p_2$. A permutation of the set of possible
 outcomes can be achieved in two manners, which may be called
 ''active'' and ''passive''. In the passive point of view the
 permutation is obtained by a simple relabelling of the possible
 outcomes and the property of invariance is self evident because
 relabelling obviously does not make an experiment more
 predictable.

 \begin{figure}
 \includegraphics[angle=0,width=0.36\textwidth]{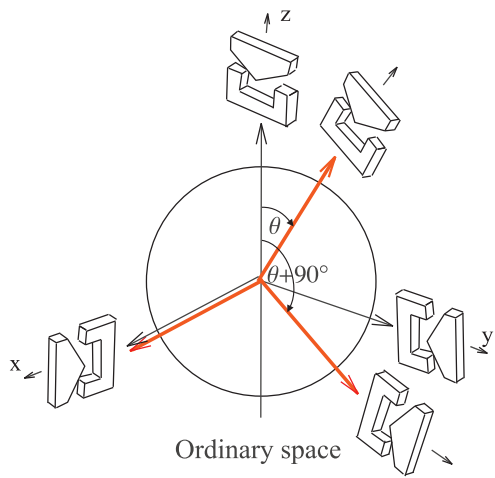} \caption{A set of
 three mutually complementary Stern-Gerlach arrangements labeled by
 a single experimental parameter $\theta$ which specifies the
 orientations of the Stern-Gerlach magnets in the three
 experiments. The three experimental arrangements are associated to
 the mutually complementary propositions: $P_1(\theta)$: ''The spin
 along the $x$-axis is up'', $P_2(\theta)$: ''The spin is up along
 the direction tilted at angle $\theta$ from the $z$-axes'' and $
 P_3(\theta)$: ''The spin is up along the direction tilted at angle
 $\theta+90^\circ$ from the $z$-axes''.} \label{rotation}
 \end{figure}

 From the active point of view, one retains the same labeling, and
 the permutation of the set of outcomes refers to a real change of
 the experimental set-up. For a spin measurement this would be a
 re-orientation of the Stern-Gerlach magnet. In that case the
 property of invariance states that the measure of information is
 indifferent under certain real physical changes of the experimental
 situation. This requirement is more stringent and may be precisely
 formulated as an invariance of the measure of information under
 interchange of the following two physical situations: (a) the
 probability for ''yes'' is $p_1$ and for ''no'' is $p_2$; and (b)
 the probability for ''yes'' is $p_2$ and for ''no'' is $p_1$. Yet
 these are different experimental situations.

 In order to remove this ambiguity in the description of the experiment
 one can assign probabilities for occurrences or different numbers
 or other distinct labels to possible outcomes, the particular scheme
 is of no further relevance. For example, one can use the statement
 "the probability for the outcome 'yes' is 0.6, and for the outcome 'no'
 is 0.4'', or the statement ''the probability for the
 outcome 'yes' is 0.4, and for the outcome 'no' is 0.6'' to
 distinguish between the situations (a) and (b) given above. Note that
in both cases the measure of information as defined by
(\ref{sunka}) is $I=0.04$.

 Here we will use a particular description which is based on the quantity
 \begin{equation}
 i=p_1-p_2. \label{latte}
 \end{equation}
 Then, on one hand, the
 sign of $i$ differs between the two situations in (a) and (b), and
 on the other hand, the square of $i$ is equal to the measure of
 information $(I=i^2)$. Therefore $i$ represents an economic and complete description
 of the experimental situation (equivalent to the
 assignment of specific probabilities for the two results)\footnote{We give another
 justification for introducing $i$. Our main goal in the
 next section will be to derive the functional dependence of
 probability $p_1(\theta)$ (recall $p_2(\theta)=1-p_1(\theta))$ on the value of the experimental parameter
 $\theta$. We will first derive the functional dependence $i(\theta)$ and
 therefrom that of $p_1(\theta)$. Note that for this purpose
 one could not use $I(\theta)$ instead of $\vec{i}(\theta)$ because
 with any value $I(\theta)$ one can associate two physically
 non-equivalent situations (a) and (b) which correspond to different values of the
 probabilities.}

 All the ''quantum state'' is meant to be is a representation of that
 catalog of our knowledge of the system that is necessary to arrive
 at the set of, in general probabilistic, predictions for all
 possible future observations of the system. Such a view was
 assumed by Schr\"{o}dinger (1935) who wrote\footnote{Translated:
 ''It (the $\psi$-function) is now the instrument for predicting the
 probability of measurement results. In it is embodied the
 respectively attained sum of theoretically grounded future
 expectations, somehow like laid down in a {\em catalogue.''}}: ''Sie
 (die $\psi$-Funktion) ist jetzt das Instrument zur Voraussage
 der Wahrscheinlichkeit von Ma{\ss}zahlen. In ihr ist die jeweils
 erreichte Summe theoretisch begr\"{u}ndeter Zukunftserwartungen
 verk\"{o}rpert, gleichsam wie in einem {\em Katalog}
 niedergelegt.'' The $\psi$ function is characterized by a set of
 complex numbers which are very remote from our everyday
 experience. Yet, if the origin of the structure of quantum
 theory is to be sought in a theory of observations, of
 observers, and of meaning, then we should focus our attention not
 on complex numbers, but rather on real-value quantities which
 are directly observable\footnote{As Peres put it: ''After all,
 quantum phenomena do not occur in a Hilbert space. They occur in a
 laboratory.''}. Interestingly, quantum theory allows descriptions of
 quantum state in terms of real numbers. An example for this is the description
 of density operators in terms of the real coefficients in the decomposition
 into generators of SU(N) algebra (basis of generalized Pauli
 matrices as used in, e.g., Schlienz and Mahler, 1998).

We will use a description of the state of an elementary system by
 a vector $\vec{i}=(i_1,i_2,i_3)=(p^+_x-p^-_x,
 p^+_y-p^-_y, p^+_z-p^-_z),$ which is a catalog of knowledge about a set of three
 mutually complementary propositions and where, in the case of spin, $p^+_x$ is the
 probability to find the particle's spin up along $x$ etc. It is assumed that the
 catalog $\vec{i}$ is a complete description of the system in the sense that its
 knowledge is sufficient to determine the probabilities for the outcomes of all possible
 future measurements.

\begin{figure}
 \includegraphics[angle=0,width=0.4\textwidth]{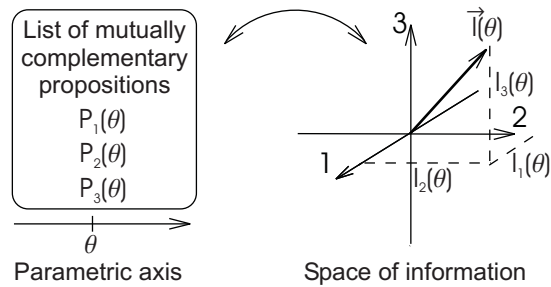}\caption{Representation of the
 state of a quantum system by the information vector
 $\vec{i}(\theta)$. The components
 $(i_1(\theta),i_2(\theta),i_3(\theta))$ of the information vector
 are associated to the three mutually complementary propositions
 $P_1(\theta), P_2(\theta)$ and $P_3(\theta)$.} \label{state}
 \end{figure}

 Denote by $\theta$ an arbitrary direction within the $y$-$z$ plane and oriented at
 an angle $\theta$ with respect to
 the $z$-axis. Now, for all $\theta$ the propositions: $P_1(\theta)$:
 ''The spin is up along the direction $x$'', $P_2(\theta)$: ''The
 spin is up along the direction $\theta$,'' and
 $P_3(\theta)$: ''The spin is up along the direction
 $\theta+90^\circ$'' are mutually complementary. The different
 lists of the three mutually complementary propositions are labeled by
 a single experimental parameter $\theta$ as given in Fig.
 \ref{rotation}. They correspond to different representations
 $\vec{i}(\theta)=(i_1(\theta),i_2(\theta),i_3(\theta))$ of the
 catalog of our knowledge of the system as shown in Fig.
 \ref{state}.

 \section{Total Information content of a quantum system}

 \label{brot}

 The finiteness of the information content of a quantum system
 comprises not just extreme cases of maximal knowledge of one
 proposition at the expense of complete ignorance of complementary
 ones but it also applies to intermediate cases. For example, it
 has been pointed out that in the interference experiments one can
 obtain some partial knowledge about the particle's path and still
 observe an interference pattern of reduced contrast as compared
 to the ideal interference situation (Wootters and Zurek, 1979;
 Englert 1999). In other words the information content of the
 system can manifest itself as path information or as modulation of the
 interference pattern or partially in both to the extent defined
 by the finiteness of information (Brukner and Zeilinger,
 2002). How to define then the total information content of a quantum
 system?

 Bohr (1958) remarked that ''... phenomena under different
 experimental conditions, must be termed complementary in the sense
 that each is well defined and that together they exhaust all
 definable knowledge about the object concerned''. This suggests
 that the total information content of a quantum system is somehow
 contained in the full set of mutually complementary experiments.
 We define the total information (of 1 bit) of the elementary (or binary,
 or two-state) system as a sum of the individual measures of information
 over a complete set of three mutually complementary experiments
 \begin{equation}
 I_{total} = I_1 + I_2 + I_3 = 1. \label{srce}
 \end{equation}

 How to define the total information content of more complex systems?
In a $n$-dimensional Hilbert space, one needs $n^2-1$ real
parameters to specify a general density matrix $\rho$, which must
be hermitean and have $Tr(\rho) = 1$.  Since measurements within
a particular basis set can yield only $n-1$ independent
probabilities (the sum of all probabilities for all possible
outcomes in an individual experiment is one), one needs $n+1$
distinct basis sets to provide the required total number of
$n^2-1$ independent probabilities. Ivanovi\'{c} (1981) showed that
the required number $n+1$ of unbiased basis sets indeed exists if
$n$ is a prime number, and Wootters and Fields (1989) showed that
it exists if $n$ is any power of a prime number.\footnote{The
composite system consisting of $N$ elementary systems with
dimension $n\!=\!2^N$ of the Hilbert space is a special case.}.
This suggests that the complete information represented by the
density matrix is fully contained in a complete set of mutually
complementary observables.

 Except for an elementary system (see Sec. \ref{NNumber}) we
 cannot give the justification for the number of mutually
 complementary observations in the general case from our basic
 considerations. We take this number in the further discussion as
 given in the quantum theory.

 Generalizing Eq. (\ref{srce}) we suggest to define the total
 information content of a $n$-dimensional quantum system as the
 sum of individual measures of information $I_{i}(\vec{p}^i)$ over
 a complete set of $n\!+\!1$ mutually complementary measurements
 \begin{equation}
 I_{total}= \sum_{i=1}^{n+1} I_{i}(\vec{p}^i) = {\cal N}
 \sum_{i=1}^{n+1} \sum_{j=1}^{n} \left( p^i_j-\frac{1}{n} \right)^2.
 \label{mast}
 \end{equation}
 Here $\vec{p^i}=(p^i_1,..., p^i_n)$ are the probabilities for
 the outcomes in the $i$-th measurement. In the case of a system
 composed of $N$ elementary systems and with appropriate normalization
it results in just $N$ bits of information (for the system in a pure state).

 The question whether or not one can find a complete set of
 mutually complementary observations in the general case of a
 Hilbert space of arbitrary dimensions is still open. If it should turn out to
 be the case, then the definition (\ref{mast}) can be applied to arbitrarily dimensional
 quantum systems. If, in contrast, such sets only exist if the dimension is the
 power of a prime number, then we suggest to take this seriously, as implying
 that the prime number alternatives are the most basic informational constituents of
 all objects (see also introduction). On the basis of this assumption the information
 content of a complex system of general dimension could be defined as a sum of the
 information contents of its individual constituents (each with the dimension of a prime
 number) plus the information contained in the correlations between them.

 For example, the system of dimension $n=p_1\cdot p_2$ where $p_1$ and $p_2$ are the prime-number
 factors of $n$ can be considered as a composite system consisting of two subsystems of
 dimensions $p_1$ and $p_2$ (The lowest dimension for which the existence of a complete set
 of mutually complementary observables has not been proven is $n\!=\!6$. There $p_1\!=\!2$ and $p_3\!=\!3$.).
 One obtains $p^2_1-1$ independent numbers from a complete set of mutually complementary measurements
 of the first subsystem and $p^2_2-1$ independent numbers from such set of measurements of the second subsystems.
 Additional $(p^2_1-1)(p^2_1-1)$ numbers can be obtained from the correlations for joint measurements
 of the two subsystems. Therefore one obtains altogether $(p^2_1\!-\!1)(p^2_1\!-\!1) \!+ \!(p^2_1\!-\!1)\!+\! (p^2_1\!-\!1)
 \!=\! p^2_1p^2_2\!-\!1$ independent parameters, which is the number of independent parameters which completely define
 the density operators of a system of dimension $p_1p_2$.

 \section{Malus law in quantum physics}

 Quantum theory predicts $p(\theta)=\cos^2(\theta/2)$ for the
 probability to find the spin up along the direction at an angle
 $\theta$ with respect to the direction along which the system
 gives spin up with certainty. From what deeper foundation emerges
 this law in quantum mechanics, originally formulated by Malus\footnote{Etienne
 Louis Malus (1775-1812), a French physicist, was almost entirely
 concerned with the study of light. He conducted experiments to
 verify Huygens' theory of light and rewrote the theory in
 analytical form. His discovery of the polarization of light by
 reflection was published in 1809 and his theory of double
 refraction of light in crystals in 1810.} for light?
 The most important contributions so far in that
 direction are those of Wootters (1981), Summhammer (1988, 1994)
 and Fivel (1994). In this section we argue that the most natural
 functional relation $p(\theta)$ consistent with the principle of quantization of
 information is indeed the sinusoidal dependence of Malus.

 We wish to specify a mapping of $\theta$ onto $\vec{i}(\theta)$.
 It is of importance to note that we can invent this mapping
 freely. The reason for this is that $\theta$ will have functional
 relations to other physical parameters of the experiment. Then,
 the laws relating those parameters with the information vector
 $\vec{i}(\theta)$ can be seen as laws about relations between those parameters
 and $\theta$ plus a mapping of $\theta$ onto $\vec{i}(\theta)$. What
 basic assumptions should we follow to obtain the mapping from
 $\theta$ to $\vec{i}(\theta)$ most appropriate for quantum mechanics?

\begin{figure}
 \includegraphics[angle=0,width=0.18\textwidth]{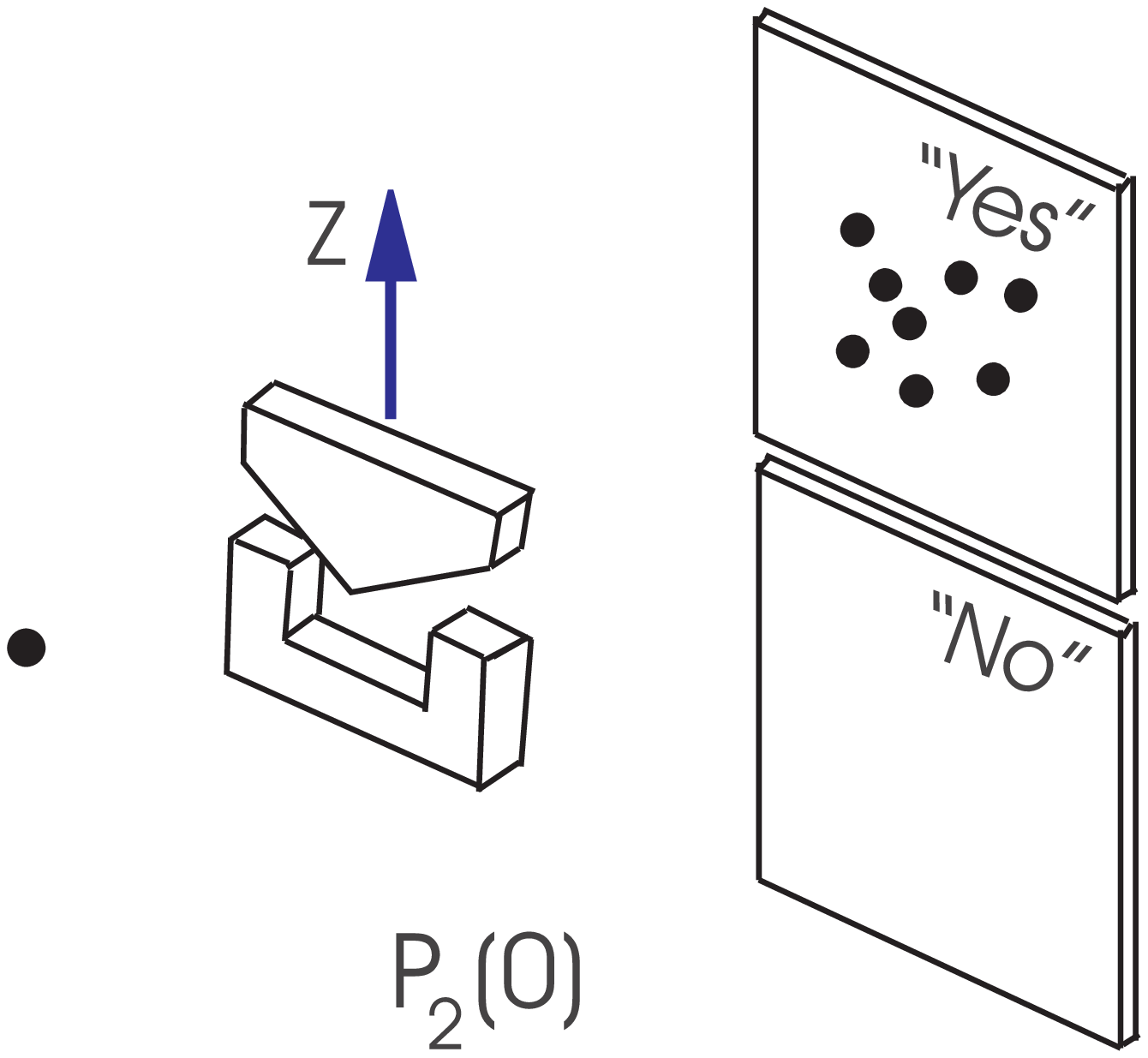} 
 \includegraphics[angle=0,width=0.26\textwidth]{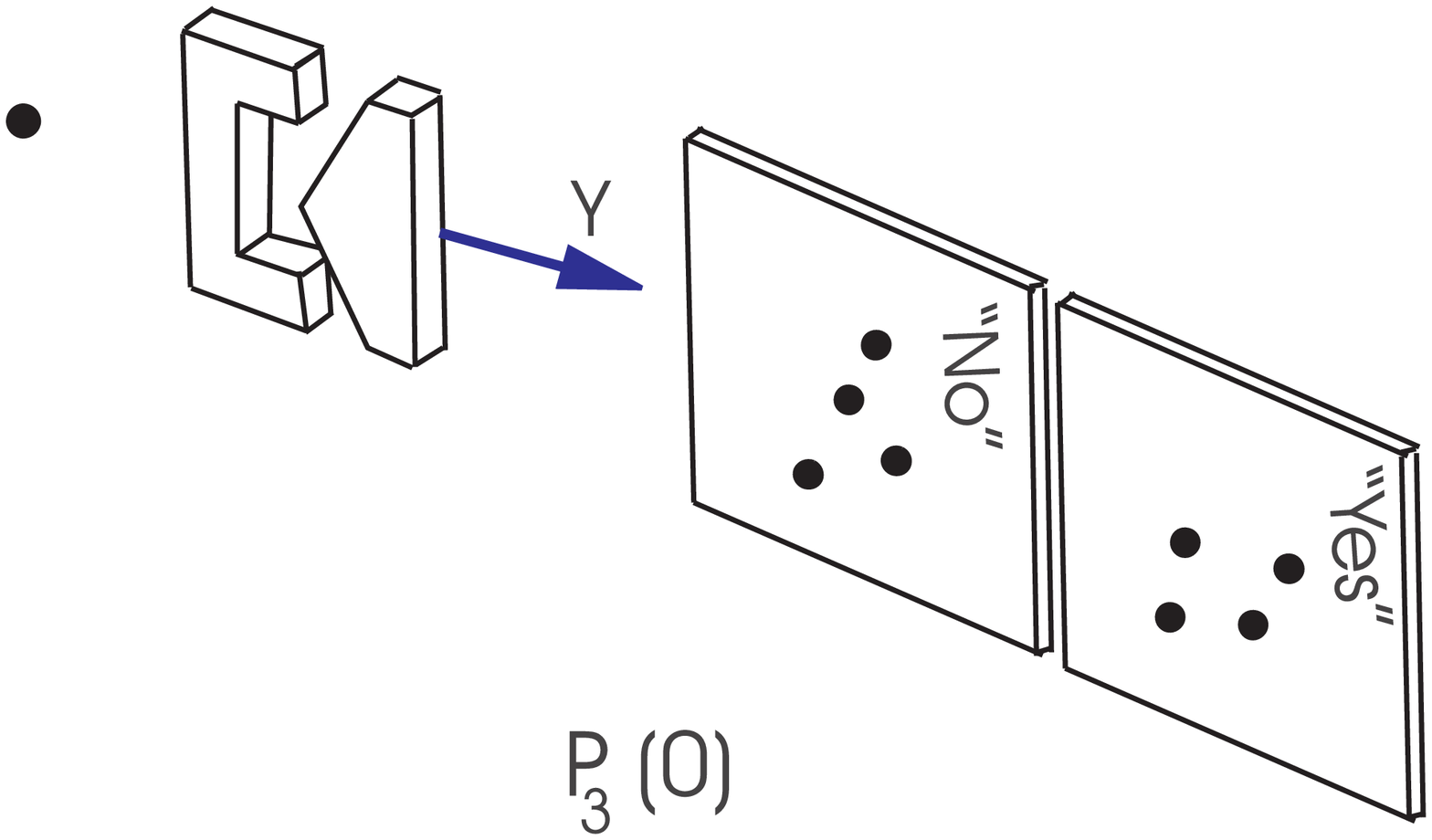}
 \hspace{1cm}
 \includegraphics[angle=0,width=0.21\textwidth]{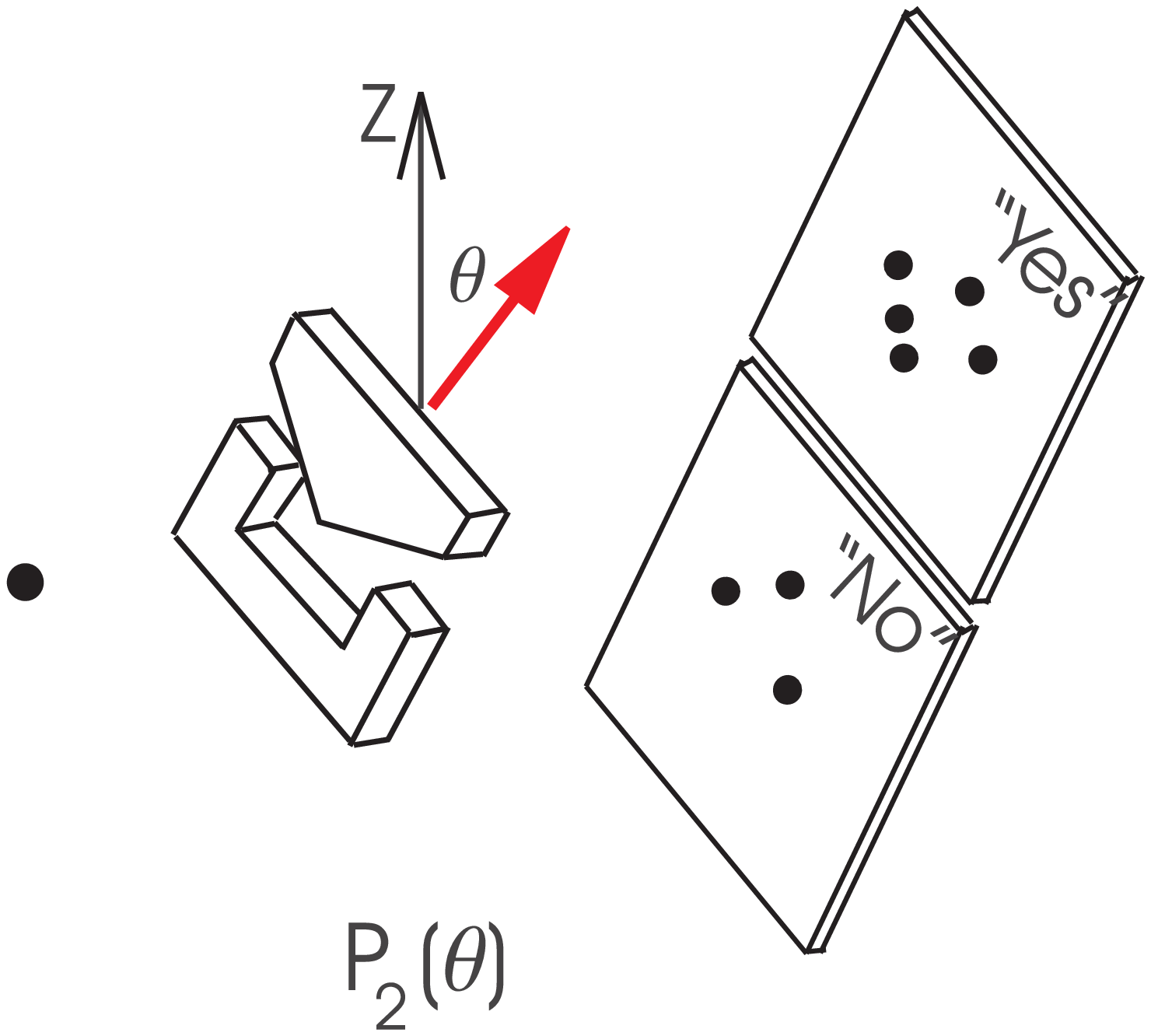} 
 \includegraphics[angle=0,width=0.23\textwidth]{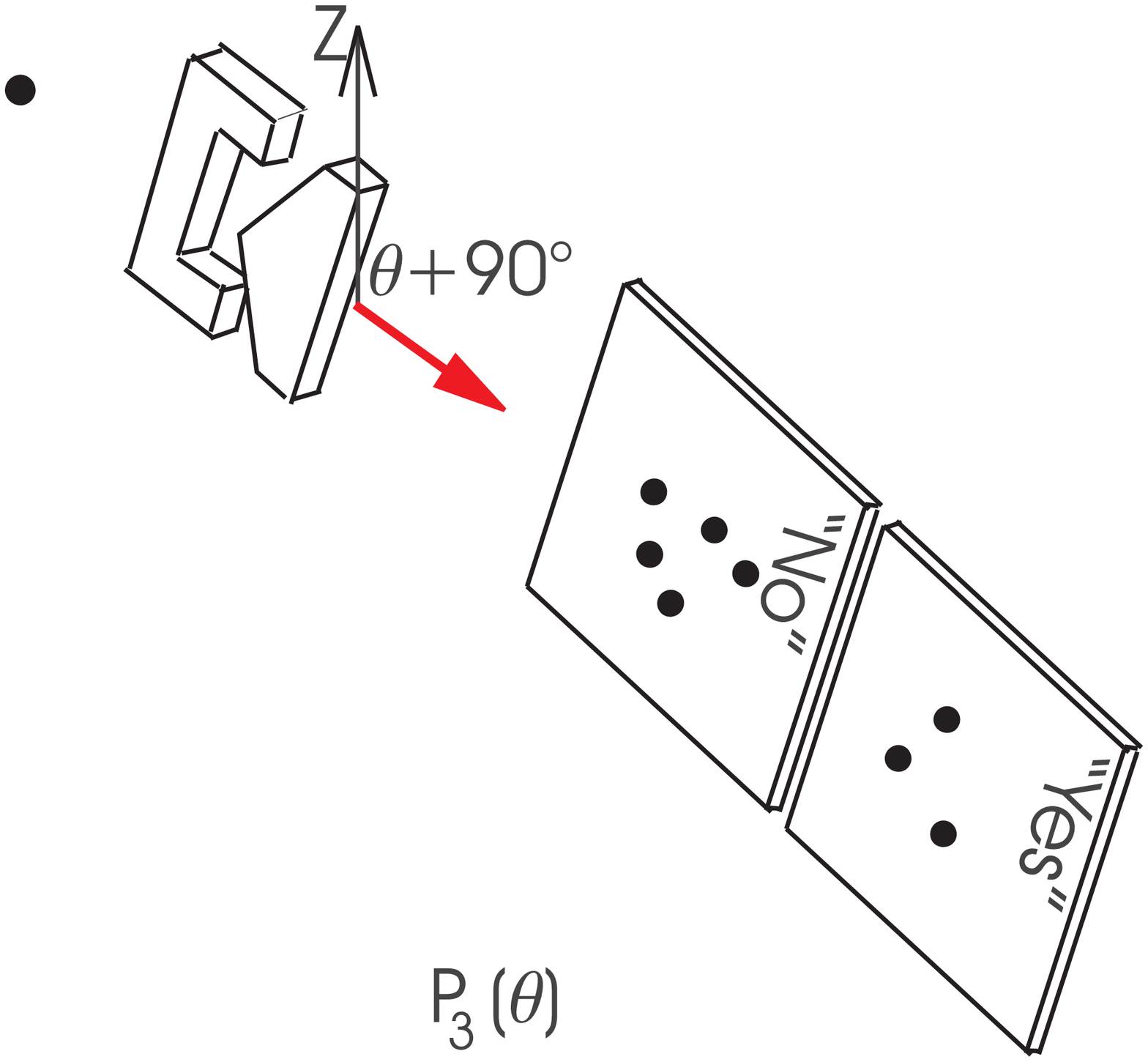} \caption{Two different sets of
 mutually complementary spin measurements (the full sets include
 also the spin measurement along the $x$-axis which is not shown in
 the figures). They correspond to the following two sets of
 mutually complementary propositions: \{$P_1(0)$: ''The spin along
 the $x$-axis is up'', $P_2(0)$: ''The spin along the $y$-axis is up'',
 $P_3(0)$: ''The spin along the $z$-axis is up''\}, and \{$
 P_1(\theta)$: ''The spin along $x$-axis is up'', $P_2(\theta)$:
 ''The spin along the direction tilted at angle $\theta$ from the
 $z$-axes is up'', $P_3(\theta)$: ''The spin along the direction
 tilted at angle $\theta+90^\circ$ from the $z$-axes is up''\}. The
 total information carried by the spin is independent of the
 particular set of mutually complementary propositions considered,
 i.e. $I_{total}\!=\!I_1(0)+I_2(0)+I_3(0)\!=\!0+0+1\!=\!
 I_1(\theta)+I_2(\theta)+I_3(\theta)$ in the example shown.}
 \label{setphoton}
 \end{figure}

 There are two basic assumptions. The first one is the assumption
 of the {\it invariance} of the total information content under the
 change of representation of the catalog of our knowledge of the
 system. Or, in other words, it is the assumption that
 total information content must be independent of the particular choice of
 mutually complementary propositions considered (see Fig.
 \ref{setphoton}). In the same spirit as choosing a coordinate
 system, one may choose any set of mutually complementary
 propositions to represent our knowledge of the system and the total
 information about the system must be invariant under that
 choice, i.e. for all $\theta$
 \begin{equation}
 I_{total} = I_1(\theta)+I_2(\theta)+I_3(\theta) =
 i^2_1(\theta)+i^2_2(\theta)+i^2_3(\theta)=1. \label{sommer}
 \end{equation}
 In fact, this property of invariance is the reason why we may use
 the phrase ''the total information content of the system ''
 without explicitly specifying a particular reference set of
 mutually complementary propositions.

\begin{figure}
 \includegraphics[angle=0,width=0.5\textwidth]{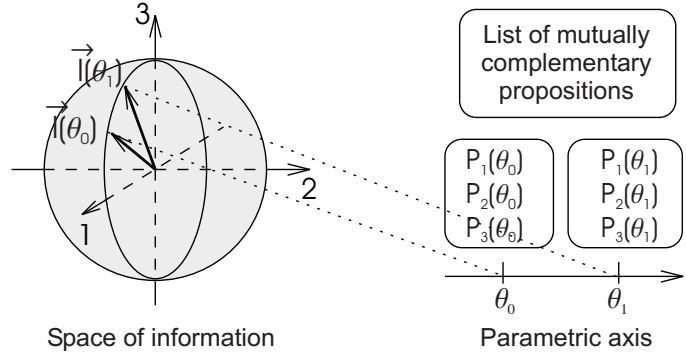}\caption{A general rotation of
 the information vector from $\vec{i}(\theta_0)$ to
 $\vec{i}(\theta_1)$ due to a change of the physical parameter from
 $\theta_0$ to  $\theta_1$.} \label{kugel}
 \end{figure}

 We suggest that only mappings where neighboring values of $\theta$
 correspond to neighboring values of $\vec{i}(\theta)$ are natural. Thus,
 if we gradually change the orientation of the magnets in a set of Stern-Gerlach
 apparata defining a complete set of mutually complementary observables a continuous
 change of the information vector will result. The property of invariance
 defined by Eq. (\ref{sommer}) implies that with a gradual
 change of the experimental parameter from $\theta_0$ to $\theta_1$ the
 information vector rotates in the space of information
 \begin{equation}
 \vec{i}(\theta_1)=\hat{R}(\theta_1-\theta_0,\theta_0)\vec{i}(\theta_0),
 \label{fm4}
 \end{equation}
 such that the length of the information vector is conserved
 (Fig. \ref{kugel}). The rotation matrix depends on two independent
 variables $\theta_0$ and $\theta_1$; here specific arguments $\theta_1\!-\!\theta_0$
 and $\theta_0$ are chosen in the functional dependence for convenience. Equation (\ref{fm4}) expresses
 our expectation that the transformation law is linear\footnote{\label{mrkva}Precisely speaking,
 the invariance property only implies that the
 transformation law $\vec{i}(\theta_1)=\vec{f}(\theta_1,\theta_0,\vec{i}(\theta_0))$
 is described by a general mapping $\vec{f}$ which preserves the length of the
 information vector. Now, consider the situation where with probability $w_A$ a
 system is prepared in state $\vec{i}_A(\theta_0)$ and with probability $w_B$ in
 $\vec{i}_B(\theta_0)$. Then the information vector is given by
 $\vec{i}(\theta_0)= w_A\vec{i}_A(\theta_0) + w_B\vec{i}_B(\theta_0)$. This means that the
 state $\vec{i}(\theta_0)$, where $\vec{i}(\theta_0)$ can, just formally, be written as
 $w_A\vec{i}_A(\theta_0) + w_B\vec{i}_B(\theta_0)$, is equivalent to the state of the system
 which is with probability $w_A$ prepared in state $\vec{i}_A$ and with
 probability $w_B$ in state $\vec{i}_B$. Let us now suppose that the experimental parameter in each of the
 three mutually complementary experiments is changed from the value
 $\theta_0$ to $\theta_1$. The individual information vectors
 $\vec{i}_A(\theta_0)$ and $\vec{i}_B(\theta_0)$ evolve
 independently, resulting in $w_A \vec{f}(\theta_1,\theta_0,
 \vec{i}_A(\theta_0)) + w_B \vec{f}(\theta_1,\theta_0,
 \vec{i}_B(\theta_0))$ for the total information vector at
 $\theta_1$. This shows that the function $\vec{f}$ is linear:
 $\vec{f}(\theta_1,\theta_0,w_A \vec{i}_A(\theta_0) + w_B
 \vec{i}_B(\theta_0))=w_A \vec{f}(\theta_1,\theta_0,
 \vec{i}_A(\theta_0)) + w_B \vec{f}(\theta_1,\theta_0,
 \vec{i}_B(\theta_0))$ for convex sums over $\vec{i}_A$ and $\vec{i}_B$. For an extension of the proof
 to arbitrary sums follow the idea from Appendix 1 of (Hardy, 2001a), which is there applied
 in a different context.}, that is, independent of the actual information vector transformed.
 $\hat R(\theta_1-\theta_0,\theta_0)$ is an orthonormal matrix
 \[
 \hat{R}^{-1}(\theta_1-\theta_0,\theta_0) =
 \hat{R}^T(\theta_1-\theta_0,\theta_0).
 \]
 Notice that transformation matrices do not build up a group in
 general because of the explicit dependence on both the initial and
 final parametric value.

\begin{figure}
\includegraphics[angle=0,width=0.46\textwidth]{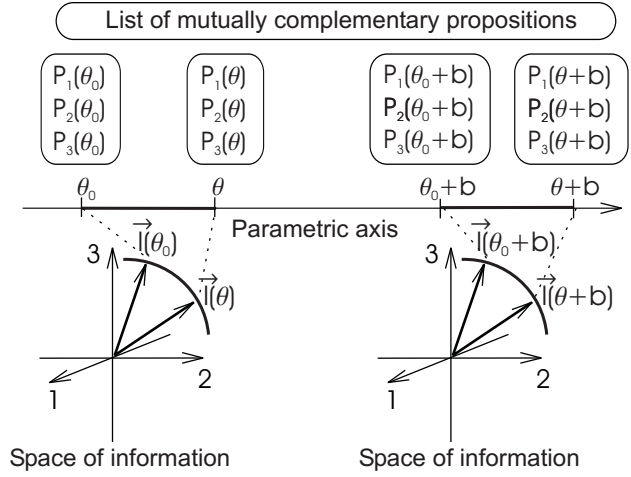} \caption{The
homogeneity of the parametric $\theta$-axis.} \label{homogenity}
\end{figure}

 The second basic assumption in the derivation of the Malus law in quantum physics
 is that no physical process {\it a priori} distinguishes one
 specific value of the physical parameter from others, that is,
 that the parametric $\theta$-axis is {\it homogeneous}. In our
 example with the orientation of Stern-Gerlach magnets as an
 experimental parameter, the homogeneity of the parametric axis
 becomes equivalent to the {\it isotropy} of the ordinary space.
 The homogeneity of the parametric axis precisely requires that if
 we transform physical situations of three complementary
 experiments together with the state of the system along the
 parametric axis for any real number $b$, we cannot observe any
 effect. Using a more formal language this means the following.
 Suppose two lists each with three mutually complementary
 experimental arrangements are associated with a specific parametric
 value $\theta_0$ and to some other value $\theta_0 + b$
 ($-\infty<b<+\infty$) respectively. Furthermore, suppose the
 information vectors $\vec{i}(\theta_0)$ and $\vec{i}(\theta_0+b)$
 associated with the two lists are equal (i.e. all components of the
 two vectors are equal). The homogeneity of the parametric
 $\theta$-axis then requires that if we change the physical
 parameter in each experiment by an equal interval of
 $\theta-\theta_0$ in the two lists of complementary experiments,
 the resulting information vectors will be equivalent as shown in
 Fig. \ref{homogenity}. Mathematically, if $\vec{i}(\theta_0) =
 \vec{i}(\theta_0+b)$  for all $\theta_0$ implies
 $\hat{R}(\theta-\theta_0,\theta_0)  \vec{i}(\theta_0)=
 \hat{R}(\theta-\theta_0,\theta_0+b)  \vec{i}(\theta_0+b)$,
 then\footnote{We give another line of reasoning, that is to
 require the {\it same} functional dependence of the transformation
 law for each initial value $\theta_0$ of the parameter. This can
 only be done with Eq. (\ref{smrad}).}
 \begin{equation}
 \hat{R}(\theta-\theta_0,\theta_0)=
 \hat{R}(\theta-\theta_0,\theta_0+b). \label{smrad}
 \end{equation}
 The transformation matrix then depends only on the difference
 between the initial and final value of the experimental parameter,
 and not on the location of these values on the parametric
 $\theta$-axis.

 The orthogonality condition leads to the following general form of
 the transformation matrix
 \begin{equation}
 \hat{R}(\theta)= \left ( \begin{array}{ccc} 1 & 0 & 0 \\ 0 &
 f(\theta) & -g(\theta) \\ 0 & g(\theta) & f(\theta)
 \end{array}     \right),
 \label{love}
 \end{equation}
 where we take $\theta_0\!=\!0$ for simplicity and
$f(\theta)$ and $g(\theta)$ are not yet specified but
 assumed to be analytical functions satisfying
 \begin{equation}
 f^2(\theta) + g^2(\theta) =1, f(0)=1 \mbox{ and } g(0)=0.
 \label{humanrace}
 \end{equation}

 We further require that a change of the experimental parameter in
 a set of mutually complementary arrangements from $\theta_0$ to
 $\theta_1$ and subsequently from $\theta_1$ to $\theta_2$ must
 have the same physical effect as a direct change of the parameter
 from $\theta_0$ to $\theta_2$. The resulting transformation will
 then be independent, whether we apply two consecutive
 transformations $\hat{R}(\theta_1-\theta_0)$ and
 $\hat{R}(\theta_2-\theta_1)$ or a single transformation
 $\hat{R}(\theta_2-\theta_0)$
 \begin{equation}
 \hat{R}(\theta_2-\theta_0) = \hat{R}(\theta_2-\theta_1)
 \hat{R}(\theta_1-\theta_0). \label{hunters}
 \end{equation}
 This together with the property that for $\theta=\theta_0$ the
 transformation matrix equals the unity matrix (since there is
 no change of the physical situations of the complementary
 experiments one has $\hat{R}(0)\!=\!\hat{1}$) implies that transformation
 matrices build up the group of rotations SO(3), a connected
 subgroup of the group of orthogonal matrices O(3) which contains
 the identity transformation.

 For the special case of infinitesimally small variation of the
 experimental conditions, Eq. (\ref{hunters}) reads
 \begin{equation}
 \hat{R}(\theta+d\theta) = \hat{R}(\theta) \hat{R}(d\theta).
 \end{equation}
 Inserting the form (\ref{love}) of the transformation matrix into
 the latter expression, one obtains
 \begin{equation}
 f(\theta+d\theta) = f(\theta)f(d\theta)-g(\theta)g(d\theta).
 \label{lola}
 \end{equation}

 Using conditions (\ref{humanrace}), we transform Eq. (\ref{lola})
 into the differential equation
 \begin{equation}
 \frac{df(\theta)}{d\theta}= - n \sqrt{1-f^2(\theta)},
 \end{equation}
 where
 \begin{equation}
 n = - g'(0) \label{mol}
 \end{equation}
 is a constant. The solution of the differential equation reads
 \begin{equation}
 f(\theta)=\cos n\theta,
 \end{equation}
 where we integrate between $0$ and $\theta$ using the condition
 $f(0)=1$ from Eq. (\ref{humanrace}). This finally leads to
 \begin{equation}
 \hat{R}(\theta)= \left ( \begin{array}{ccc} 1 & 0 & 0 \\ 0 & \cos
 n\theta  & -\sin n\theta \\ 0 & \sin n\theta & \cos n\theta
 \end{array}     \right).
 \label{dayin}
 \end{equation}
 This result directly gives the familiar expression
 \begin{equation}
 p=\cos^2 \frac{n \theta}{2} \label{cosinus}
 \end{equation}
 for probability in quantum theory.

Mathematically one could consider our result as a direct and
immediate consequence of the theory of group representations; the
cosine dependence follows from a particular representation of the
rotation group. However
 from the physical perspective it is implied by fundamental assumptions:
 (1) the total information of the system is invariant under the
 change of the representation of the catalog of our knowledge about
 the system and (2) the parametric space is homogeneous. If (1) and (2) are satisfied, then the
 probability must vary as $\cos^2 n\theta$, where $n$ is a parameter not determined by the
 derivation. Quantum-mechanical probabilities are just of this form, with $\theta$ for a relative
 polarization angle and with $n=1/2$ for electrons and neutrinos,
 or with $n=1$ for photons, or with $n=2$ for gravitons. The same
 functional dependence $\cos^2 \phi$ undergoes also the probability
 to find a particle in a specific output beam in the Mach-Zehnder
 type of interferometer with the phase shift $\phi$ between two
 paths inside the interferometer.

 In the discussion so far we considered a change of a single
 experimental parameter and the rotation of the information vector
 within one plane only. This can be generalized. Let us
 define the orientations of the three mutually orthogonal
 directions $\vec{n}_1(\alpha,\beta,\gamma)$,
 $\vec{n}_2(\alpha,\beta,\gamma)$ and
 $\vec{n}_3(\alpha,\beta,\gamma)$ in ordinary space by the Euler
 angles $0 \!\leq \!\alpha \!<\! 2\pi$, $0\! \leq \!\beta \!\leq \!\pi$, and $0 \!\leq\!
 \gamma \!< \!2\pi$. Then the mutually complementary propositions
 which are associated to measurements along the three directions
 can be represented in terms of the Euler angles as
 $P_1(\alpha,\beta,\gamma)$: ''The spin along the direction
 $\vec{n}_1(\alpha,\beta,\gamma)$ is up,''
 $P_2(\alpha,\beta,\gamma)$: ''The spin along the direction
 $\vec{n}_2(\alpha,\beta,\gamma)$ is up'' and
 $P_3(\alpha,\beta,\gamma)$: ''The spin along the direction
 $\vec{n}_3(\alpha,\beta,\gamma)$ is up''.

 Given a specific set of three orthogonal directions, all other
 sets of orthogonal directions can be obtained by rotating the
 reference set. Any general rotation for Euler's angles
 $\alpha,\beta,\gamma$ can be performed as a sequence of three
 rotations, the first around the $z$-axes by $0\leq\gamma<2\pi$, the
 second around the new $y$-axes by $0\leq\beta\leq\pi$ and finally
 the third around the new $z$-axes by $0\leq\alpha<2\pi$.

 A list of mutually complementary propositions associated to the
 spin measurements along directions obtained by the first rotation
 is $P_1(0,0,\gamma)$, $P_2(0,0,\gamma)$, and $P_3(0,0,\gamma)$. Following the argumentation given above one
 obtains
 \begin{equation}
 \hat{R}(\gamma)= \left ( \begin{array}{ccc} \cos\gamma &
 -\sin\gamma & 0 \\ \sin\gamma  & \cos\gamma  & 0 \\ 0 & 0  &
 1
 \end{array}     \right)
 \label{evolution}
 \end{equation}
 for the corresponding transformation matrix in the space of
 information\footnote{One should always keep in mind the difference
 between directions along which mutually complementary measurements
 are performed in ordinary space (such as the vertical direction
 and the direction at $+45^\circ$ along which a photon's polarization is
 measured, or three spatially orthogonal directions along which complementary
 spin components of a spin-1/2 particle are measured) and directions associated with mutually
 complementary propositions (components of an information vector)
 in the space of information. The latter always constitute an
 orthogonal coordinate system. These again have to be distinguished from the
 orthogonal directions in Hilbert space which do not correspond to complementary
 measurements.}. If we fix the angle of the first
 rotation at $\gamma_0$ and consider only propositions $P_1(0,\beta,\gamma_0)$, $P_2(0,\beta,\gamma_0)$ and $ P_3(0,\beta,\gamma_0)$ about spins along directions obtained by
 the second rotation around the new $y$-axis for an angle
 $0\!\leq\!\beta\!\leq\!\pi$, the corresponding transformation matrix reads
 \begin{equation}
 \hat{R}(\beta)= \left ( \begin{array}{ccc} \cos\beta & 0 &
 \sin\beta \\ 0 & 1 & 0 \\ -\sin\beta & 0 & \cos\beta
 \end{array}     \right).
 \end{equation}
 In the last step we fix both the angle $\gamma_0$ of the first
 rotation and the angle $\beta_0$ of the second rotation, and
 consider only sets of mutually complementary propositions
 $P_1(\alpha,\beta_0,\gamma_0)$, $P_2(\alpha,\beta_0,\gamma_0)$
 and $P_3(\alpha,\beta_0,\gamma_0)$ about spins along directions
 obtained by the third rotation around the new $z$-axis for
 $0\!\leq\!\alpha\!<\!2\pi$. The corresponding transformation matrix is
 again of the form (\ref{evolution}) with the angle $\alpha$.

 Finally, the transformation matrix for a general rotation in the
 space of information is given as
 \begin{equation}
 \hat{R}(\alpha,\beta,\gamma)=\hat{R}(\alpha)\hat{R}(\beta)\hat{R}(\gamma).
 \end{equation}

 While these relations have an obvious meaningful for spin they hold equally for
 any elementary system. Specifically they also hold for a two-path interferometer.

 \section{Entanglement - More information in joint properties than in
 individuals}

 Entanglement is the feature which distinguishes quantum physics
 most succinctly from classical physics as quantitatively expressed by the
 violation of Bell's inequalities (Bell 1964, Clauser
 {\it et al.}, 1969). In 1964 John Bell obtained
 certain bounds (the Bell inequalities) on combinations of statistical
 correlations for measurements on two-particle systems if these
 correlations were to be understood within a realistic picture based on
 local properties of each individual particle. In a
 such a picture the measurement results are determined by
 properties the particles carry prior to and independent of
 observation. In a local picture the results obtained at one
 location are independent of any measurements or actions performed
 at space-like separation. Quantum mechanics predicts violation of
 these constraints for certain statistical predictions for the
 composite (entangled) systems. By today, the predictions of quantum
 physics have been confirmed in many experiments (Freedman and Clauser 1972;
 Aspect {\it et al.}, 1981; Weihs {\it et al.}, 1998; Pan {\it et al.}, 2000)

 In this section we will investigate how much information can be
 contained in the correlations between quantum systems in order to
 give an information-theoretic criterion of quantum entanglement.
 We suggest that a natural understanding of quantum entanglement
 results when one accepts that the information in a composite
 system can reside more in the correlations than in properties of
 individuals. The quantitative formulation of these ideas leads to
 a rather natural criterion of quantum entanglement\footnote{To
 this end we will follow Schr{\"o}dinger's (1935) view about
 entanglement: ''Whenever one has a complete expectation-catalog -
 a maximum total knowledge - a psi-function - for two completely
 separated bodies, or, in better terms, for each of them singly,
 then one obviously has it also for the two bodies together, i.e.,
 if one imagines that neither of them singly but rather the two of
 them together make up the object of interest, of our questions about the future.
But the converse is not true. Maximal knowledge of a total system
does not necessarily include total knowledge of all its parts,
not even when these are fully separated from each other and at
the moment are not influencing each other at all.''}.

 The total information of a composite system can be distributed in
 various ways within the composite system. We will consider only that
 part of the total information of the system which is exclusively
 contained in correlations, or joint properties of its constituents. This is
 also the reason why now we will not consider complete sets of
 mutually complementary propositions for the composite system but
 just that subset of them which concerns joint properties of its
 constituents. As it is our final goal to compare that criterion
 with the one given by Bell-type inequalities where one considers
 correlations between spin
 measurements confined on each side within one plane we restrict
 our analysis to an $x$-$y$ plane locally defined for each
 subsystem.

 As an explicit example of a composite systems a system consisting of
 two spin-1/2 particles will be considered. The propositions about their
 joint properties will be binary propositions, i.e. will be associated
 to experiments with two possible outcomes. The two outcomes will
 correspond to the proposition of the type: "The spin of particle 1 along
 $x$ and the spin of particle 2 along $y$ are the same", and
 to its negation "The spin of particle 1 along $x$ and the spin
 of particle 2 along $y$ are different''. Therefore the measure of
 information Eq. (\ref{sunka}) for binary experiments can be applied. If we denote the
 probabilities for the two outcomes by $p^{+}_{xy}$ and
 $p^{-}_{xy}$ respectively, then the information contained
 in proposition "The spin of particle 1 along
 $x$ and the spin of particle 2 along $y$ are the same (different)" is given
 by
 \begin{equation}
 I_{xy}=(p^{+}_{xy}- p^{-}_{xy})^2.
 \label{measure2}
 \end{equation}

 We first consider a product state e.g. $|\psi\rangle=|+x\rangle_1|-x\rangle_2$.
 This is the case of a composite system composed of two
 elementary systems carrying therefore $N=2$ bits of information,
 i.e. representing the truth value of two propositions.
 Here the state  $|\psi\rangle$ represents the two-bit combination true-false
 of the truth values of the propositions about the spin of each
 particle along the $x$-axis: (1) "The spin of particle 1 is up
 along $x$" and (2) "The spin of particle 2 is up along $x$".
 Instead of the second proposition describing the spin of particle
 2, we could alternatively choose a proposition which describes the
 result of a joint observation: (3) "The two spins are the same
 along $x$." Then the state $|\psi\rangle$ represents the two-bit
 combination true-false of the truth values of the propositions (1)
 and (3).

 Evidently, for pure product states at most one proposition with
 definite truth-value can be made about joint properties because
 one proposition has to be used up to define a property of one of
 the two subsystems. In other words 1 bit of information defines
 the correlations. In our example where $|\psi\rangle=|\!+x\rangle_1|\!-x\rangle_2$ the
 correlations are fully represented by the correlations between
 spin $x$-measurements on the two sides, therefore
 \begin{equation}
 I_{xx}=1.
 \label{prodcorr}
 \end{equation}
 We denote the states with property (\ref{prodcorr}) as classically composed
 states.

 Obviously, the choice of directions $x$ and $y$ within each of the planes of
 measurements on the two sides is arbitrary. It is physically not acceptable
 that the total information contained in correlations between spin
 measurements confined on each side within $x$-$y$ planes depends on this choice.
 We therefore require that the total information contained in the correlations
 must be invariant upon the choice of general $x$ and $y$ measurement directions
 within the $x$-$y$ planes on each side. Only with this requirement the statement
 "the total information contained in the correlations between measurements within the
 $x$-$y$ planes" can have a meaning independent of the specific
 set of mutually complementary measurements considered. This
 invariance property can only be guaranteed with our measure of
 information (\ref{measure2}).

 We define the total information contained in the correlations as the sum over the individual
 measures of information about a complete set of mutually complementary observations within the planes $x$-$y$
 on the two sides. The total information contained in the correlations is thus defined as
 the sum
 \begin{equation}
 I_{corr}=I_{xx}+I_{xy}+I_{yx}+I_{yy} \label{corr}
 \end{equation}
 of the partial measures of information contained in the set of
 complementary observations within the $x$-$y$-planes. These
 observations are mutually complementary for product states and the set is complete as there exists no
 further complementary observation within the chosen $x$-$y$
 planes. By this we mean that for any product state a complete
 knowledge contained in any proposition from the set: "The two spins are
 equal along $x$", "The spin of particle 1 along $x$ and the spin
 of particle 2 along $y$ are the same", "The spin of particle 1
 along $y$ and the spin of particle 2 along $x$ are the same" and
 "The two spins are equal along $y$" excludes any knowledge about
 other three propositions.

In general there can also be some amount of information contained
in the correlations for measurements involving $z$ direction, for
example, for measurement directions within the $x$-$z$ planes on
the two sides. Obviously if the general $x$ and $y$ directions
are chosen to include directions outside of the old $x$-$y$ planes
measure of information $I_{corr}$ cannot be assumed to remain an
invariant. The maximal value of $I_{corr}$ can then be obtained by
an optimization over all possible two-dimensional planes of
measurements on both sides.

 Consider now a maximally entangled Bell state, e.g.
 \begin{eqnarray}
 |\psi^-\rangle&=&\frac{1}{\sqrt{2}}(|+x\rangle_1|-x\rangle_2-|-x\rangle_1|+x\rangle_2)
 \nonumber
 \\
 &=&\frac{1}{\sqrt{2}}(|+y\rangle_1|-y\rangle_2-|-y\rangle_1|+y\rangle_2).
 \label{bellstate}
 \end{eqnarray}
 The two propositions here both are statements about results of
 joint observations (Zeilinger 1997), namely (1') "The two spins
 are equal along $x$" and (2') "The two spins are equal
 along $y$". Now the state represents the two-bit combination
 false-false of these propositions. Note that
 here the 2 bits of information are all carried by the 2 elementary
 systems in a joint way, with no individual elementary system
 carrying any information on its own. In other words, as the two
 available bits of information are already exhausted in defining
 joint properties, no further possibility exists to also encode
 information in individuals. Therefore
 \begin{equation}
 I^{Bell}_{corr}=2. \label{bellcorr}
 \end{equation}
 Note that in our example of Bell-state (\ref{bellstate}) $I_{xx}=I_{yy}=1$  and $I_{xy}=I_{yx}=0$.
 Also, note that the truth value for another proposition, namely,
 "The two spins are equal along $z$" must follow immediately from
 the truth values of the propositions (1') and (2'), as only 2 bits
 of information are available. Interestingly this is also a direct
 consequence of the formalism of quantum mechanics as the joint
 eigenstate of   $\sigma^1_x \sigma^2_x$ and of   $\sigma^1_y
 \sigma^2_y$  is also an eigenstate of $\sigma^1_z \sigma^2_z =
 -(\sigma^1_x \sigma^2_x) (\sigma^1_y \sigma^2_y)$.

 In contrast to product states we suggest entanglement of two
 elementary systems to be defined in general such that {\it more
 than one bit} (of the two available ones) is used to define joint
 properties, i.e.
 \begin{equation}
 I^{entgl}_{corr}>1 \label{entanglcorr}
 \end{equation}
 for at least one choice of the planes of measurements for
 the two elementary systems (or, equivalently, for that choice of the
 planes of measurements for which $I_{corr}$ reaches
 its maximal value). Most importantly this simple information-theoretic
 criterion of entanglement can be shown to be  equivalent to a {\it necessary}
 and {\it sufficient} condition (Horodeccy family, 1995) for a violation of a Bell-type
 inequality for two-elementary systems. A generalization of our information-theoretic
 criterion for entanglement to $N$ elementary systems and its relation to the criteria
for violation of Bell's inequalities can be found in (Brukner {\it
et al.}, 2001).

 \section{Time Evolution of the Catalog of Knowledge}

 \label{DDynamics}

 Any assignment of properties to an object is always a consequence of some observation.
 Using information obtained in previous observations
 we wish to make predictions about the future. Again our
 predictions might be formulated as, in general probabilistic,
 predictions about future properties of a system. Clearly, these
 predictions can be verified or falsified by performing
 measurements and checking whether the experimental results agree
 with our predictions. It is then important to connect past
 observations with future observations. Or, more precisely, to make, based on past
 observations, specific statements about possible results of future observations.

 In quantum mechanics this connection between past observation and
 future observation exactly is achieved by the quantum-mechanical
 Liouville equation (for pure states it reduces to the
 Schr\"{o}dinger equation)
 \begin{equation}
 i\hbar\frac{d\hat{\rho}(t)}{dt}=[\hat{H}(t),\hat{\rho}(t)].
 \end{equation}

 The initial state $\hat{\rho}(t_0)$ represents all our information
 as obtained by earlier observation. Using the quantum-mechanical
 Liouville equation we can derive a time evolved final state
 $\hat{\rho}(t)$ at some future time $t$ which gives us predictions
 for any possible observation of the system at that time. In this section the
 dynamics of an elementary system is formulated as a time evolution
 of the catalog of our knowledge of the system. This is specified
 by the evolution of the information vector in the space of
 information. The Liouville equation will then be derived from the
 differential equation describing the motion of the information
 vector in the information space.

 We will consider now the time evolution of an elementary system with
 no information exchange with an environment\footnote{If there is information
 exchange between the system and the environment we cannot
 formulate system's evolution law independently of the environment,
 but we have to consider it as a subsystem of a larger system that
 contains both the system and the environment where again the total information
 is conserved.}. Suppose that the state of the system at some initial time $t_0$
 is represented by the catalog $\vec{i}(t_0)=(i_1(t_0), i_2(t_0),
 i_3(t_0))$ of our knowledge. Now let the system evolve in time.
 Because there is no information exchange with an environment
 during the evolution, the total information of the system at some
 later time $t$ must still be the same as at the initial time. This
 may be seen as an ultimate constant of the evolution of the system
 motion independent of the strength, time dependence or any other
 characteristic of the "external field" of the system. Therefore
 \begin{equation}
 I_{total}(t)= \sum_{n=1}^{3} i^2_n(t) = \sum_{n=1}^{3} i^2_n(t_0)
 = I_{total}(t_0).
 \end{equation}

 Mathematically, the conservation of the total information is
 equivalent to the conservation of the length of the information
 vector during its motion in the information space. This means that time
 evolution of an isolated quantum system is just a rotation of
 the information vector in the space of information (see footnote \ref{mrkva})
 \begin{equation}
 \vec{i}(t)=\hat{R}(t,t_0)\vec{i}(t_0), \label{fm5}
 \end{equation}
 where again $\hat{R}(t,t_0)$ is a rotation matrix
 \[
 \hat{R}^{-1}(t,t_0) = \hat{R}^T(t,t_0)
 \]
and $\hat{R}^T(t,t_0)$ is its transposed matrix.

 The derivative of Eq. (\ref{fm5}) with respect to time is
 \begin{equation}
 \frac{d\vec{i}(t)}{dt} = \frac{d\hat{R}(t,t_0)}{dt}\vec{i}(t_0)=
 \hat{K}(t,t_0)\vec{i}(t), \label{lifeisover}
 \end{equation}
 where
 $\hat{K}(t,t_0)=\frac{d\hat{R}(t,t_0)}{dt}\hat{R}^{T}(t,t_0)$. We
 will now show that the operator $\hat{K}(t,t_0)$ is
 antisymmetric. We find
 \begin{eqnarray*}
 \hat{K}^{T}(t) &=& \hat{R}(t)\frac{d\hat{R}^{T}(t)}{dt} =
 \hat{R}(t) \lim_{\Delta t \rightarrow 0}
 \frac{\hat{R}^{T}(t+\Delta t)-\hat{R}^{T}(t)} {\Delta t} =
 \hat{R}(t) \lim_{\Delta t \rightarrow 0} \hat{R}^{T}(t)
 \frac{\hat{R}(t)-\hat{R}(t+\Delta t)}{\Delta t}
 \hat{R}^{T}(t+\Delta t) \\ &=& \lim_{\Delta t \rightarrow 0}
 \frac{\hat{R}(t)-\hat{R}(t+\Delta t)} {\Delta t} \hat{R}^{T}(t) =
 -\hat{K}(t),
 \end{eqnarray*}
 where the initial time $t_0$ is identified with the time 0.

 It is a well-known result of vector analysis that with every
 antisymmetric operator $\hat{K}$ one may uniquely associate the
 ''vector of rotation'' $\vec{u}$ by the relation\footnote{The
 operator $\hat{K}$ is represented by an antisymmetric matrix
 \[
 \hat{K}= \left ( \begin{array}{ccc} 0 & -k_{21} & -k_{31} \\
 k_{21} & 0 & -k_{32} \\ k_{31} & k_{32} & 0
 \end{array}     \right).
 \]
 From there we read out the components of the vector of rotation
 $\vec{u}$ as
 \[
 u_{1}=k_{32}, u_{2}=-k_{31}, u_{3}=k_{21}.
 \]}
 \begin{equation}
 \hat{K}\vec{y}=\vec{u}\times \vec{y} \hspace{1cm} \mbox{ for all
 } \vec{y},
 \end{equation}
 where ''$\times$'' denotes vector product. Using this result we
 now rewrite Eq. (\ref{lifeisover}) as
 \begin{equation}
 \frac{d\vec{i}(t)}{dt}=\vec{u}(t,t_0) \times \vec{i}(t).
 \label{please}
 \end{equation}
 Mathematically, this equation describes the rotation of the
 information vector around the axis $\vec{u}(t,t_0)$ which itself
 changes in the course of time. Physically, this is the formulation of the
 dynamical law for the evolution of the catalog of our knowledge.

One might recognize Eq. (\ref{please}) as a description of the
state evolution in terms of the Bloch vector. Based on the known
features of the quantum formalism we will now argue for the
validity of Eq. (\ref{please}). Suppose that the quantum
 state of the system is described by the density matrix
 $\hat{\rho}$. We decompose the density matrix into the unity
 operator and the generators of SU(2) algebra (Pauli matrices)
 \begin{equation}
 \hat{\rho}(t)=\frac{1}{2}\hat{1} + \frac{1}{2}\sum_{j=1}^3
 i_j(t)\hat{\sigma}_j, \label{smisao}
 \end{equation}
 where $\hat{\sigma}_j$ is spin operator for the direction $j=x,y,z$.
 Note that the quantity $i_j$ for the spin along the
 direction $j$ is equal to the expectation value of spin along this
 direction, i.e. $i_j(t)=\mbox{Tr}(\hat{\rho}(t)\hat{\sigma}_j)$.

 If we take a derivative of Eq. (\ref{smisao}) in time we obtain
 \begin{equation}
 i\hbar\frac{d\hat{\rho}(t)}{dt}=\frac{1}{2}\sum_{j=1}^3
 \frac{i_j(t)}{dt}\hat{\sigma}_j.
 \end{equation}
 Inserting Eq. (\ref{please}) on the right-hand side we find
 \begin{equation}
 i\hbar\frac{d\hat{\rho}(t)}{dt}=\frac{i}{2}\sum_{i,j,k=1}^3\epsilon_{ijk}
 u_i(t) i_j \hat{\sigma}_k.
 \end{equation}
 Since the Pauli matrices satisfy
 $[\hat{\sigma}_i,\hat{\sigma}_j]=2i\sum_{k=1}^3
 \epsilon_{ijk}\hat{\sigma}_k$, we proceed with
 \begin{equation}
 i\hbar\frac{d\hat{\rho}(t)}{dt}=\frac{1}{4} \sum_{i,j=1}^3 u_i(t)
 i_j (\hat{\sigma}_i \hat{\sigma}_j-\hat{\sigma}_j \hat{\sigma}_i).
 \end{equation}
 Introducing the operator $\hat{H}(t)$ such that
 \begin{equation}
 u_{i}(t) := \mbox{Tr}(\hat{H}(t) \hat{\sigma}_i), \label{hamiltonian}
 \end{equation}
 we finally obtain the quantum-mechanical Lioville equation
 \begin{equation}
 i\hbar\frac{d\hat{\rho}(t)}{dt}=[\hat{H}(t),\hat{\rho}(t)].
 \end{equation}

\begin{figure}
 \includegraphics[angle=0,width=0.54\textwidth]{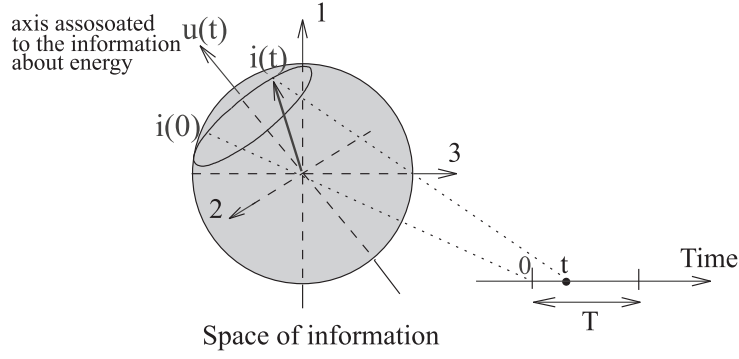} \caption{One
 complete rotation of the information vector after a time elapse of
 the de-Broglie wave-period.} \label{frequency}
 \end{figure}

 For the special case of a conservative system, the evolution of a
 quantum state in time is constrained by a higher constant of
 motion, namely our information about the energy of the system,
 apart from the ultimate one of the total information content of
 the system. In the space of information this corresponds to the
 rotation of the information vector around a fixed axis that is
 associated to our knowledge of energy of the system\footnote{Note
 that we consider elementary systems, that is, systems with two
 possible energy values. By information about the energy of the
 system we mean our knowledge about which of the two values will be
 observed in an appropriately designed experiment.}. This is only
 possible if the axis $\vec{u}$ in Eq. (\ref{please}) is a fixed
 axis in time around which the information vector rotates. This
 further implies the existence of a minimal interval of time the
 information vector needs to make one complete rotation in the
 space of information (Fig. \ref{frequency}). After this time
 interval the values $i$ for all propositions about the system take
 the same value. This time interval is known as the deBroglie
 wave-period.

 Obviously the result given above is just a very first and the simplest step
 toward an information-theoretical formulation of the quantum-mechanical
 evolution in time. Mathematically, it is an immediate
 consequence of the (nearly) isomorphism between SU(2) group of
 unitary rotations in a two-dimensional Hilbert space and SO(3)
 group of rotations in the real three-dimensional Euclidian space.
 Obviously one needs to consider more complex systems and to find position-
 and momentum-representations of the catalog of our knowledge in
 order to give an information-theoretical formulation of the
 Schr{\"o}dinger equation.

 \section{Measurement - the update of information}

 \label{itfrombit}

 \hyphenation{Real-dings}

 In this section, it will be argued that identifying the quantum
 state of a system with the catalog of our knowledge of the system
 leads to the resolution of many of the seemingly paradoxical
 features of quantum mechanics connected to the so-called measurement problem.

 In a quantum measurement, we find the system to be in one of the
 eigenstates of the observable defined by the measurement
 apparatus. A specific example is the case when we are considering
 a wave packet as being composed of a superposition of plane waves.
 Such a wave packet is more or less well-localized, but we can
 always perform a position measurement on a wave packet which is
 better localized than the dimension of the packet itself. This,
 sometimes called ''reduction of the wave packet'' or ''collapse of
 the wave function'',  can only be seen as a ''measurement
 paradox'' if one views this change of the quantum state as a real
 physical process. In the extreme case it is often even related to
 an instant collapse of some physical wave in space.

 There is no basis for any such assumption. In contrast, there is
 never a paradox if we realize that the wave function is just an
 encoded mathematical representation of our knowledge of the
 system. When the state of a quantum system has a non-zero value at
 some position in space at some particular time, it does not mean
 that the system is physically present at that point, but only that
 our knowledge (or lack of knowledge) of the system allows the
 particle the {\em possibility} of being present at that point at
 that instant.

 What can be more natural than to change the representation of our
 knowledge if we gain new knowledge from a measurement performed
 on the system? When a measurement is performed, our knowledge of
 the system changes, and therefore its representation, the quantum
 state, also changes. In agreement with the new knowledge, it
 instantaneously changes all its components, even those which
 describe our knowledge in the regions of space quite distant from
 the site of the measurement. Then no need whatsoever arises to
 allude to notions like superluminal or instantaneous transmission
 of information.

 Schr\"{o}dinger (1935) wrote\footnote{Translated: ''For each measurement one
 is required to ascribe to the $\psi$-function (=the prediction
 catalog) a characteristic, quite sudden change, which {\em
 depends on the measurement result obtained}, and so {\em cannot
 be foreseen}; from which alone it is  already quite clear that
 this second kind of change of the $\psi$-function has nothing
 whatever in common with its orderly development {\em between} two
 measurements. The abrupt change by measurement ... is the most
 interesting point of the entire theory. It is precisely {\em the}
 point that demands the break with naive realism. For {\em this}
 reason one {\em cannot} put the $\psi$-function directly in place
 of the model or of the physical thing. And indeed not because one
 might never dare impute abrupt unforseen changes to a physical
 thing or to a model, but because in the realism point of view
 observation is a natural process like any other and cannot per se
 bring about an interruption of the orderly flow of natural
 events.''}: ''Bei jeder Messung ist man gen\"{o}tigt, der
 $\psi$-Funktion (=dem Voraussagenkatalog) eine eigenartige, etwas
 pl\"{o}tzliche Ver\"{a}nderung zuzuschreiben, die von der {\it
 gefundenen Ma{\ss}zahl} abh\"{a}ngt und sich {\it nicht
 vorhersehen l\"{a}{\ss}t}; woraus allein schon deutlich ist,
 da{\ss} diese zweite Art von Ver\"{a}nderung der $\psi$-Funktion
 mit ihrem regelm\"{a}ssigen Abrollen {\it zwischen} zwei
 Messungen nicht das mindeste zu tun hat. Die abrupte
 Ver\"{a}nderung durch die Messung ... ist der interessanteste
 Punkt der ganzen Theorie. Es ist genau {\it der} Punkt, der den
 Bruch mit dem naiven Realismus verlangt. Aus {\it diesem} Grund
 kann man die $\psi$-Funktion {\it nicht} direkt an die Stelle des
 Modells oder des Realdings setzen. Und zwar nicht etwa weil man
 einem Realding oder einem Modell nicht abrupte unvorhergesehene
 \"{A}nderung zumuten d\"{u}rfte, sondern weil vom realistischen
 Standpunkt die Beobachtung ein Naturvorgang ist wie jeder andere
 und nicht per se eine Unterbrechung des regelm\"{a}ssigen
 Naturlaufs hervorrufen darf''.

 A closely related position was assumed also by Heisenberg, who
 wrote in a letter to Renninger dated February 2, 1960: ''The act
 of recording, on the other hand, which leads to the reduction of
 the state, is not a physical, but rather, so to say, a
 mathematical process. With the sudden change of our knowledge also
 the mathematical presentation of our knowledge undergoes of course
 a sudden change.'', as translated by Jammer (1974).

 We will now bring the role of the observer in a quantum
 measurement to the center of our discussion. In classical physics
 we can assume that an observation reveals some property already
 existing in the outside world. For example, if we look at the
 moon, we just find out where it is and it is certainly safe to
 assume that the property of the moon to be there is independent of
 whether anyone looks or not. The situation is drastically
 different in quantum mechanics and it is just the very attitude of
 the Copenhagen interpretation giving a fundamental role to
 observation which is a major intellectual step forward over this
 naive classical realism. With the only exception of the system
 being in an eigenstate of the measured observable, a quantum
 measurement changes the system into one of the possible new states
 defined by the measurement apparatus in a fundamentally
 unpredictable way, and thus cannot be claimed to reveal a property
 existing before the measurement is performed. The reason for this
 is again the fact that a quantum system cannot, not even in
 principle, carry enough information to specify
 observation-independent properties corresponding to all possible
 measurements. In the measurement the state therefore must appear
 to be changed in accord with the new information, if any, {\it
 acquired} about the system together with unavoidable and
 irrecoverable {\it loss} of complementary information. Unlike a
 classical measurement, a quantum measurement thus does not
 just add (if any) some knowledge, it changes our knowledge in agreement
 with a fundamental finiteness of the total information content of the
 system\footnote{Wheeler (1989) stated that ''... yes or no that is
 recorded constitutes an unsplittable bit of information''.}

 We as observers have a significant role in the measurement
 process, because we can decide by choosing the measuring device
 which attribute will be realized in the actual
 measurement\footnote{Wheeler explicates this by example of the
 well-known case of a quasar, of which we can see two pictures
 through the gravity lens action of a galaxy that lies between the
 quasar and ourselves. By choosing which instrument to use for
 observing the light coming from that quasar, we can decide here
 and now whether the quantum phenomenon in which the photons take
 part is interference of amplitudes passing on both sides of the
 galaxy or whether we determine the path the photon took on one or
 the other side of the galaxy.}. Since the information content of
 the system is limited, by choosing which measurement device to use
 we not only decide what particular knowledge will be gained, but
 simultaneously what complementary knowledge will be lost after the
 measurement is performed. Here, a very subtle position was assumed
 by Pauli (1955) who writes: ''The gain of knowledge by means of an
 observation has as a necessary and natural consequence, the loss
 of some other knowledge. The observer has however the free choice,
 corresponding to two mutually exclusive experimental arrangements,
 of determining {\it what} particular knowledge is gained and what
 other knowledge is lost (complementary pairs of opposites).
 Therefore every irrevocable interference by an observation about a
 system alters its state, and creates a new phenomenon in Bohr's
 sense.''

 \section{Conclusions}

 The laws we discover about Nature do not already exist as ''Laws
 of Nature'' in the outside world. Rather ''Laws of Nature'' are
 necessities of the mind for any possibility to make sense
 whatsoever out of the data of experience. This epistemological
 structure is a necessity behind the form of all laws an observer
 can discover. As von Weizs\"acker has put it, and Heisenberg
 quoted in (1958) paper: ''Nature is earlier than man, but man is
 earlier than natural science.''

 An observer is inescapably suspended in the situation of
 obtaining the data from observation, formatting concepts of Nature
 therefrom, and predicting the data of future observations.
 In observing she/he is able to distinguish only a finite number of
 results at each interval of time (compare Summhammer, 2000; 2001).
 Therefore the experience of the ultimate experimenter is a
 stream of (''yes'' or ''no'') answers to the questions posed to
 Nature. Any concept of an existing reality is then a mental
 construction based on these answers. Of course this does not imply
 that reality is no more than a pure subjective human construct.
 From our observations we are able to build up objects with a set
 of properties that do not change under variations of modes of
 observation or description. These are ''invariants'' with respect
 to these variations. Predictions based on any such specific
 invariants may then be checked by anyone, and as a result we may
 arrive at an intersubjective agreement about the model, thus
 lending a sense of independent reality to the mentally constructed
 objects.

 In quantum experiments an observer may decide to measure a
 different set of complementary variables, thus gaining certainty
 about one or more variable at the expense of losing certainty
 about the other(s). Thus the measure of information in an individual
 experiment is not an invariant but depends on the specific
 experimental context. However the total uncertainty, or
 equivalently, the total information, is invariant under such
 transformation from one complete set of complementary variables
 to another. In classical
 physics a property of a system is a primary concept prior to and
 independent of observation and information is a secondary concept
 which measures our ignorance about properties of the system. In
 contrast in quantum physics the notion of the total information
 of the system emerges as a primary concept, independent of the
 particular complete set of complementary experimental procedures
 the observer might choose, and a property becomes a secondary
 concept, a specific representation of the information of the
 system that is created spontaneously in the measurement itself.
 Bohr (1934) wrote that '' ... a subsequent measurement to a
 certain degree deprives the information given by a previous
 measurement of its significance for predicting the future course
 of phenomena. Obviously, these facts not only set a limit to the
 {\em extent} of the information obtainable by measurement, but
 they also set a limit to the {\em meaning} which we may attribute
 to such information.''

 Theorems like those of Bell (1964) and Greenberger-Horne-Zeilinger
 (1990) state that randomness of an individual quantum
 event cannot be derived from local causes (local
 hidden variables). Quantum physics is not able to ''explain why
 (specific) events happen'' as pointed out by Bell (1990). It is
 beyond the scope of quantum physics to answer the question why
 events happen at all (that is, why the detectors clicks at
 all). Yet, if events happen, then they must happen randomly.
 The reason is the finiteness of the information.
 Any detailed description of the reality
 that would be able to give an unambiguous answer to Bell's
 question, that is, any description that would be able to arrive
 at an accurate and detailed prediction of the particular process
 resulting in a particular event, will necessarily include the
 definition of a number of ''hidden'' properties of the system
 which would carry information as to which specific result will be
 observed for all possible future measurements. Therefore no answer
 can be given to Bell's question, because otherwise, quantum system would
 carry more information that it is in principle available.

 It turns out that the lowest symmetry common for
 all elementary systems is the invariance of their total information
 content with respect to a rotation in a three-dimensional space.
 The three dimensionality of the information space is a
 consequence of the minimal number (3) of mutually exclusive
 experimental questions we may pose to an elementary system. This
 seems to justify the use of three-dimensional space as ''the''
 space of the inferred world. Such a view was first suggested by v.
 Weizs\"{a}cker (1974): "It [quantum theory of the simple
 alternative] contains a two-dimensional complex vector space with
 a unitary metric, a two-dimensional Hilbert space. This theory
 has a group of transformations which is surprisingly
 near-isomorphic with a group of rotations in the real
 three-dimensional Euclidian space. This has been known for a very
 long time. I propose to take this isomorphism seriously as being
 the real reason why ordinary space is three-dimensional."

 We end with another quote of v. Weizs\"{a}cker (1974): "But I feel
 these consideration make it plausible that quantum theory is not
 just one out of a thousand equally possible theories, and the one
 which happens to please God so much that he chose to create a
 world in which it would be true. I rather think, if we had
 understood quantum theory just a little bit better than we
 understand it so far it would turn out to be a fairly good
 approximation towards the formulation of a theory which contains
 nothing but the rules under which we speak about future events if
 we can speak about them in an empirically testable way at all."

 It has not escape our attention that our considerations presented
 here may be viewed as providing the necessary justification for this point of
 view.

 \section*{Acknowledgements} We acknowledge discussions with Terry Rudolph,
 Christoph Simon, Johann Summhammer and Marek {\. Z}ukowski. This work
 is supported by the Austrian  FWF project F1506, and by the QIPC
 program of the EU.

\section{REFERENCES}

\hspace{-0.47cm}  Aspect, A., P. Grangier, and G. Roger, 1981,
Phys. Rev. Lett.
{\bf 47}, 460-463. \\
Bell, J. S., 1964, Physics 1, 195-200; reprinted Bell, J. S.,
1987, {\it Speakable and Unspeakable in Quantum Mechanics}
(Cambridge Univ. Press).\\
Bell, J. S., 1990, Physics World (August 1990). \\
Bohr N., 1949, in {\it Albert Einstein: Philosopher-Scientist},
edited by P.A. Schillp (The Library of Living Philosophers
Evanston, IL) 200. A copy can be found at the web site
(http://www.emr.hibu.no/lars/eng/schlipp/Default.html).\\
Bohr, N., 1958, {\it Atomic Physics and Human Knowledge} (Wiley,
New York). \\
Brukner, \v C., and A. Zeilinger, 1999, Phys. Rev. Lett. {\bf 83},
3354-3357. \\
Brukner, \v C. and A. Zeilinger, 2000, e-print quant-ph/0008091.
\\
Brukner, \v C, M. \.Zukowski, and A. Zeilinger, 2001, e-print
quant-ph/0106119.\\
Brukner, \v C. and A. Zeilinger, 2001, Phys. Rev. A {\bf 63},
022113 1-10.\\
Brukner, \v C. and A. Zeilinger, 2002, Phil. Trans. R. Soc. Lond.
A 360 (2002) 1061.\\
Caves, C. M., C. A. Fuchs, R. Schack, 2001a, e-print
quant-ph/0104088.\\
Caves, C. M., C. A. Fuchs, R. Schack 2001b, Phys. Rev. A 022305.\\
Clauser, J., M. Horne, A. Shimony, and R. Holt, 1969, Phys. Rev.
Lett. {\bf 23}, 880-884.\\
Donath, N., and K. Svozil, 2002, Phys. Rev. A {\bf 66}, 044302. \\
Freedman, S. J., and J. S. Clauser, 1972, Phys. Rev. Lett. {\bf
28}, 938-941.\\
Englert B. G., 1996, Phys. Rev. Lett. {\bf 77}, 2154.\\
Fivel, D.I., 1994, Phys. Rev. A {\bf 59}, 2108.\\
Fuchs, C. A., 2002, e-print quant-ph/0205039.\\
Fuchs, C. A., 2001, e-print quant-ph/0106166.\\
Gnedenko, B. V., 1976, {\it The Theory of Probability} (Mir
Publishers, Moscow).\\
Greenberger, D. M., M. Horne, A. Shimony, and A. Zeilinger, 1990,
Am. J. Phys. {\bf 58}, 1131-1143.\\
Hall, M. J. W., 2000, e-print quant-ph/0007116.\\
Hardy G., J. E. Littlewood and G. P\'{o}lya, 1952 {\it
Inequalities} (Cambridge University Press).\\
Hardy, L., 2001a, e-print quant-ph/0101012.\\
Hardy, L., 2001b, e-print quant-ph/0111068.\\
Heisenberg, W., 1958, Daedalus {\bf 87}, 95.\\
Horodecki, R., P. Horodecki, and M. Horodecki, 1995, Phys. Lett. A
{\bf 200}, 340-344.\\
Ivanovi{\'c} I., 1981, J. Phys. A {\bf 14}, 3241.\\
Jammer, M., 1974, {\em The Philosophy of Quantum Mechanics}, (J.
Wiley \& Sons, New York).\\
Kochen, S. and E. P. Specker, 1967, J. Math. and Mech. {\bf 17},
59.\\
Landauer R., 1991 May, Physics Today, 23.\\
Pan, J. W., D. Bouwmeester, H. Weinfurter, and A. Zeilinger,
2000, Nature {\bf 403}, 515-518.\\
Pauli W., 1955, in {\it Writings on Philosophy and Physics}
edited by C. P. Enz and K. von Meyenn, translated by Robert
Schlapp (Springer Verlag, Berlin).\\
Schlienz J. and G. Mahler, 1995, Phys. Rev. A {\bf 52},
4396-4404.\\
Shannon, C. E., 1948, Bell Syst. Tech. J. {\bf 27}, 379. A copy
can be found at
(http://cm.bell-labs.com/cm/ms/what/shannonday/paper.html).\\
Schr\"{o}dinger, E., 1935, Naturwissenschaften 23, 807-812;
823-828; 844-849. Translation published in {\it Proc. Am. Phil.
Soc.} 124, p. 323-338 and in {\it Quantum Theory and Measurement}
edited by J. A. Wheeler and W. H. Zurek (Princeton University
Press, Princeton), p. 152-167. A copy can be found at
(www.emr.hibu.no/lars/eng/cat).\\
Summhammer, J., 1988, Found. Phys. Lett. {\bf 1}, 123.\\
Summhammer, J., 1994, Int. J. Theor. Phys. {\bf 33}, 171.\\
Summhammer, J., 2000, e-print quant-ph/0008098, to appear in "The
Third Millenium" edited by Cristian Calude.\\
Summhammer, J. 2001, e-print quant-ph/0102099.\\
Svozil, K., 2002, Phys. Rev. A {\bf 66}, 044306. \\
Timpson, C. G., 2001, e-print quant-ph/0112178.\\
von Weizs\"{a}cker, C. F., 1958, {\em Aufbau der Physik} (Carl
Hanser,
M\"unchen).\\
von Weizs\"{a}cker, C. F., 1974, in {\it Quantum Theory and the
Structures of Time and Space}, edited by L. Castell, M.
Drieschner, C. F. von Weizs\"{a}cker (Hanser, M\"{u}nchen, 1975).
Papers presented at a conference held in Feldafing, July 1974.\\
Weihs, G., T. Jennewein, C. Simon, H. Weinfurter, and A.
Zeilinger, 1998, Phys. Rev. Lett. {\bf 81}, 5039-5043.\\
Wheeler J. A., 1983, Law without Law in {\em Quantum Theory and
Measurement} edited by J. A. Wheeler and W. H. Zurek (Princeton
University Press, Princeton) 182.\\
Wheeler J. A., 1989, Proc. 3rd Int. Symp. Foundations of Quantum
Mechanics, Tokyo, 354.\\
Wootters, W. K., 1981, Phys. Rev D {\bf 23}, 357.\\
Wootters W. K. and B. D. Fields, 1989, Ann. Phys. {\bf 191},
363.\\
Wootters, W. K., and W. H. Zurek, 1979, Phys. Rev. D {\bf 19},
473.\\
Zeilinger, A., 1997, Phil. Trans. Roy. Soc. Lond. {\bf 1733},
2401-2404.\\
Zeilinger, A., 1999, Found. Phys. {\bf 29}, 631-643.\\


\end{document}